\documentstyle[epsf]{livrev}
\begin{document}

\title{Quasi-Normal Modes of Stars and Black Holes}

\author{
Kostas D. Kokkotas \\
Department of Physics,
Aristotle University of Thessaloniki,\\
Thessaloniki 54006, Greece.\\
kokkotas@astro.auth.gr\\
http://www.astro.auth.gr/\~{}kokkotas \\
\\
\\
and
\\
\\
Bernd G. Schmidt\\
Max Planck Institute for Gravitational Physics,\\ 
Albert Einstein Institute, D-14476 Golm, Germany.\\
bernd@aei-potsdam.mpg.de\\
\\
\\
{\em (Accepted 31 August 1999)}
}


\date{}
\maketitle

\centerline{\bf Abstract}
Perturbations of stars and black holes have been one of the main topics
of relativistic astrophysics for the last few decades. 
They are of particular importance today, because of their relevance 
to gravitational wave astronomy. 
In this review we present the theory of quasi-normal modes of       
compact objects from both the mathematical and astrophysical points of view. 
The discussion includes perturbations of black holes        
(Schwarzschild, Reissner-Nordstr\"om, Kerr and  Kerr-Newman) 
and relativistic stars (non-rotating and slowly-rotating). 
The properties of the various families of quasi-normal modes are described, 
and numerical techniques for calculating quasi-normal modes reviewed. 
The successes, as well as  the limits, of perturbation theory are 
presented, and its role in the emerging era of numerical relativity 
and supercomputers is discussed.\\      


\newpage

\section{Introduction}
\label{section_1}

Helioseismology and asteroseismology are well known terms in 
classical astrophysics. From the beginning of the century the
variability of Cepheids has been used for the accurate measurement
of cosmic distances, while the variability of a number of 
stellar objects (RR Lyrae, Mira) has been associated with stellar
oscillations. Observations of solar oscillations (with thousands of
nonradial modes) have also revealed a wealth of information about the
internal structure of the Sun~\cite{Unno}. Practically every stellar
object oscillates radially or nonradially, and although there is great
difficulty in observing such oscillations there are already results
for various types of stars (O, B, \dots). All these types of
pulsations of normal main sequence stars can be studied via  Newtonian
theory and they are of no importance for the forthcoming era of
gravitational wave astronomy. The gravitational waves emitted by these stars are
extremely weak and have very low frequencies (cf. for a
discussion of the sun ~\cite{CL96}, and an important new measurement of the sun's
quadrupole moment and its application in the
measurement of the anomalous precession of Mercury's perihelion ~\cite{Pijpers}). 
This is not the case when we consider very compact stellar objects
i.e.\ neutron stars and black holes. Their oscillations, produced mainly
during the formation phase, can be strong enough to be detected by the
gravitational wave detectors (LIGO, VIRGO, GEO600, SPHERE) which are
under construction. 

In the framework of general relativity (GR) quasi-normal modes (QNM) 
arise, as perturbations (electromagnetic or
gravitational) of stellar or black hole spacetimes. Due to the emission
of gravitational waves there are no normal mode oscillations but
instead the frequencies become ``quasi-normal'' (complex), with the
real part representing the actual frequency of the oscillation and the
imaginary part representing the damping.

In this review we shall discuss the oscillations of neutron stars and 
black holes. The natural way to study these oscillations is by
considering the linearized Einstein equations. Nevertheless, there has
been  recent work on  nonlinear black hole
perturbations~\cite{GNPP96, GNPP96b, GNPP98a, GNPP98b, GNPP98c} while, 
as yet nothing is known for nonlinear stellar oscillations in general
relativity.
 
The study of black hole perturbations was initiated by the pioneering
work of Regge and Wheeler~\cite{RW57} in the late 50s and was
continued by Zerilli~\cite{Zer70}. The perturbations of relativistic
stars in GR were first studied in the late 60s by Kip Thorne and
his collaborators~\cite{TC67, Thorne68, Thorne69a, Thorne69b}. The
initial aim of Regge and Wheeler was to study the stability of a black
hole to small perturbations and they did not try to connect these
perturbations to astrophysics. In contrast, for the case of
relativistic stars, Thorne's aim was to extend the known properties of
Newtonian oscillation theory to general relativity, and to estimate
the frequencies and the  energy radiated as  gravitational waves.

\begin{figure}[hptb]   
  \epsfysize=6.5cm  
  \epsfxsize=8.5cm 
  \centerline{\epsffile{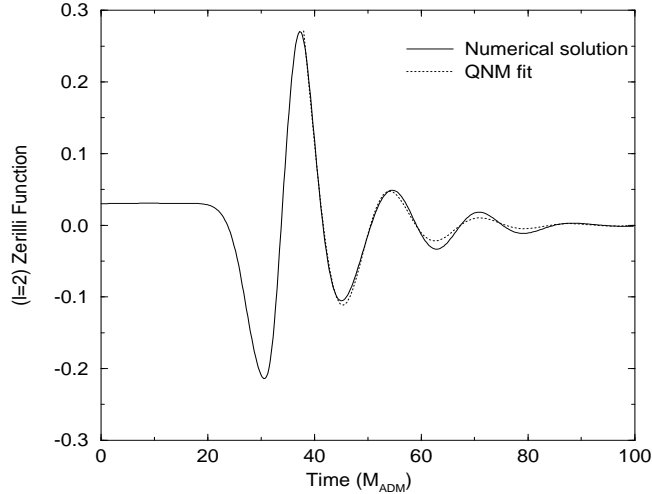}}  
  \caption{\it QNM ringing after the head-on collision of two unequal mas    s black holes~\cite{AB98}. The continuous line corresponds to the full
    nonlinear numerical calculation while the dotted line is a fit to
    the fundamental and first overtone QNM.}
  \label{qnm1}   
\end{figure}

QNMs were first pointed out by Vishveshwara~\cite{vishu} in
calculations of the scattering of gravitational waves by a
Schwarzschild black hole, while Press~\cite{press71} coined the term
{\em quasi-normal frequencies}. QNM oscillations have been found  in
perturbation calculations of particles falling into
Schwarzschild~\cite{DRPP71} and Kerr black holes~\cite{Det77, DS79}
and in the collapse of a star to form a black hole~\cite{CPMa, CPMb,
CPMc}. Numerical investigations of the fully nonlinear equations of
general  relativity have provided results which agree with the results
of perturbation calculations; in particular numerical studies
of the head-on collision of two black holes~\cite{AHSSS93, AB98}
(cf. Figure~\ref{qnm1}) and gravitational collapse to a Kerr
hole~\cite{SP85}. Recently, Price, Pullin and
collaborators~\cite{PP94, APPSS95, GNPP96, AP97} have pushed forward
the agreement between full nonlinear numerical results and results from
perturbation theory for the collision  of two black holes. This proves
the power of the perturbation approach even in highly nonlinear
problems while at the same time indicating its limits.

In the concluding remarks of their pioneering paper on nonradial
oscillations of neutron stars Thorne and Campollataro~\cite{TC67}
described it as ``{\em just a modest introduction to a story which
promises to be long, complicated and fascinating}''. The story has
undoubtedly proved to be intriguing, and many authors have contributed
to our present understanding of the pulsations of both black holes and
neutron stars. Thirty years after these prophetic words by Thorne and
Campollataro hundreds of papers have been written in an attempt to
understand the stability, the characteristic frequencies and  the
mechanisms of excitation of these oscillations. Their
relevance to the emission of gravitational waves was always the basic
underlying reason of each study. An account of all this work will be
attempted in the next sections hoping that the interested reader will
find this review useful both as a guide to the literature and as an
inspiration for future work on the open problems of the field.
In the next section we attempt to give a mathematical definition
of QNMs. The third and fourth section will be devoted to the study of
the black hole and stellar QNMs. In the fifth section we discuss the
excitation and observation of QNMs and finally in the sixth section we
will mention the more significant numerical techniques used in the
study of QNMs.


\newpage

\section{Normal Modes -- Quasi-Normal Modes -- Resonances}
\label{section_2}

Before discussing quasi-normal modes it is useful to remember what
normal modes are!

Compact classical linear oscillating systems such as finite strings, 
membranes, or cavities filled with electromagnetic radiation have
preferred time harmonic states of motion ($\omega$ is real):
\begin{equation}
  \chi_n(t,x)=e^{i\omega_n t}\chi_n(x), \quad n = 1, 2, 3 \dots,
  \label{bs1.1}
\end{equation}
if dissipation is neglected. (We assume $\chi$ to be some complex
valued field.) There is generally an infinite collection of such
periodic solutions, and the ``general solution'' can be expressed as a
superposition,
\begin{equation}
  \chi(t,x)=\sum_{n=1}^\infty a_n e^{i\omega_n t}\chi_n(x),
  \label{bs1.2}
\end{equation}
of such normal modes. The simplest example is a string of length $L$
which is fixed at its ends. All such systems can be described by
systems of partial differential equations of the type ($\chi$ may be
a vector)
\begin{equation}
  {\partial\chi\over\partial t} = {\bf A}\chi,
  \label{bs1.3}
\end{equation}
where ${\bf A}$ is a linear operator acting only on the spatial
variables. Because of the finiteness of the system the time evolution
is only determined if some boundary conditions are prescribed. 
The search for solutions periodic in time leads to a boundary value
problem in the spatial variables. In simple cases it is of the
Sturm-Liouville type. The treatment of such boundary value problems
for differential equations played an important role in the development
of Hilbert space techniques. 

A Hilbert space is chosen such that the differential operator
becomes symmetric. Due to the boundary conditions dictated by the
physical problem, ${\bf A}$ becomes a self-adjoint operator on the
appropriate Hilbert space and has a pure point spectrum. The
eigenfunctions and eigenvalues determine the periodic
solutions~(\ref{bs1.1}). 

The definition of self-adjointness is rather subtle from a physicist's
point of view since fairly complicated ``domain issues'' play an
essential role. (See~\cite{BS95} where a mathematical exposition for
physicists is given.) The wave equation modeling the finite string has
solutions of various degrees of differentiability. To describe all
``realistic situations'', clearly $C^\infty$ functions should be
sufficient. Sometimes it may, however, also be convenient to consider
more general solutions. 

From the mathematical point of view the collection of all smooth
functions is not a natural setting to study the wave equation because 
sequences of solutions exist which converge to non-smooth solutions. 
To establish such powerful statements like~(\ref{bs1.2}) one has to
study the equation on certain subsets of the Hilbert space of square
integrable functions. For ``nice'' equations it usually  happens that
the eigenfunctions are in fact analytic. They can then be used to
generate, for example, all smooth solutions by a pointwise converging
series~(\ref{bs1.2}). The key point is that we need some mathematical
sophistication to obtain the ``completeness property'' of the
eigenfunctions.  

This picture of ``normal modes''  changes when we consider ``open
systems'' which can lose energy to infinity. The simplest case are
waves on an infinite string. The general solution of this problem is
\begin{equation}
  \chi(t,x)=A(t-x) + B(t+x)
  \label{bs1.4}
\end{equation}
with ``arbitrary'' functions $A$ and $B$. Which solutions should we
study? Since we have all solutions, this is not a serious question. 
In more general cases, however, in which the general solution is not 
known, we have to select a certain class of solutions which we
consider as relevant for the physical problem.

Let us consider for the following discussion, as an example, a wave
equation with a potential on the real line,
\begin{equation}
  {\partial^2\over{\partial t^2}}\chi
  +\left(-{\partial^2\over{\partial x^2}} + V(x)\right)\chi = 0.
  \label{bs1.5}
\end{equation}
Cauchy data $\chi(0,x), \partial_t\chi(0,x)$ which have two
derivatives determine a unique twice differentiable solution. 
No boundary condition is needed at infinity to determine the
time evolution of the data! This can be established by fairly simple
PDE theory~\cite{John}.

There exist solutions for which the support of the fields are 
spatially compact, or -- the other extreme -- solutions with infinite
total energy for which the fields grow at spatial infinity in a quite
arbitrary way! 

From the point of view of physics smooth solutions with spatially
compact support should be the relevant class -- who cares what happens
near infinity! Again it turns out that mathematically it is more
convenient to study all solutions of finite total energy. Then the
relevant operator is again self-adjoint, but now its spectrum is
purely ``continuous''. There are no eigenfunctions which are square
integrable. Only ``improper eigenfunctions'' like plane waves exist.  
This expresses the fact that we find a solution of the form~(\ref{bs1.1})  
for any real $\omega$ and by forming appropriate superpositions one
can construct solutions which are ``almost eigenfunctions''. (In the
case $V(x)\equiv 0$ these are wave packets formed from plane waves.) 
These solutions are the analogs of normal modes for infinite systems.

Let us now turn to the discussion of ``quasi-normal modes'' which are
conceptually different to normal modes. To define quasi-normal modes
let us consider the wave equation~(\ref{bs1.5}) for  potentials with
$V\geq 0$ which vanish for $|x|>x_0$. Then in this case all solutions
determined by data of compact support are bounded: $|\chi(t,x)|< C$. 
We can use Laplace transformation techniques to represent such
solutions. The Laplace transform $\hat{\chi}(s,x)$ ($s>0$ real) of a
solution $\chi(t,x)$ is 
\begin{equation}
  \hat{\chi}(s,x)=\int_0^\infty e^{-st} \chi(t,x) dt,
  \label{bs1.6}
\end{equation}
and satisfies the ordinary differential equation
\begin{equation}
  s^2\hat{\chi} -\hat{\chi}''+ V \hat{\chi} = + s \chi(0,x)+\partial_t
  \chi (0,x),
  \label{bs1.7}
\end{equation}
where
\begin{equation}
  s^2\hat{\chi} -\hat{\chi}''+ V \hat{\chi} = 0
  \label{bs1.7b}
\end{equation}
is the homogeneous equation. The boundedness of $\chi$ implies that
$\hat{\chi}$ is analytic for positive, real $s$, and has an analytic
continuation onto the complex half plane $Re(s)>0$.

Which solution $\hat{\chi}$ of this inhomogeneous equation gives the
unique solution in spacetime determined by the data? There is no
arbitrariness; only one of the Green functions for the inhomogeneous
equation is correct!

All Green functions can be constructed by the following well known
method. Choose any two linearly independent solutions of the
homogeneous equation $f_-(s,x)$ and $f_+(s,x)$, and define
\begin{equation}
  G(s,x,x') = {1 \over W(s)}
  \begin{array}{ll}
    f_-(s,x') f_+(s,x) \quad & (x' < x), \\
    f_-(s,x) f_+(s,x') \quad & (x' > x),
  \end{array}
  \label{bs1.8}
\end{equation}
where $W(s)$ is the Wronskian of $f_-$ and $f_+$. If we denote the
inhomogeneity of~(\ref{bs1.7}) by $j$, a solution of~(\ref{bs1.7}) is
\begin{equation} 
  \hat{\chi}(s,x) = \int_{-\infty}^\infty G(s,x,x') j(s,x') dx'. 
  \label{bs1.9}
\end{equation}
We still have to select a unique pair of solutions $f_-, f_+$. Here
the information that the solution in spacetime is bounded can be
used. The definition of the Laplace transform implies that
$\hat{\chi}$ is bounded as a function of $x$. Because the potential
$V$ vanishes for $|x|>x_0$, the solutions of the homogeneous
equation~(\ref{bs1.7b}) for $|x|>x_0$ are
\begin{equation}
  f = e^{\pm sx}.
  \label{bs1.10}
\end{equation}
The following pair of solutions
\begin{equation}
  f_+=e^{-sx} \quad \mbox{for } x>x_0 , \quad \quad f_-=e^{+sx} \quad
  \mbox{for } x<-x_0,
  \label{bs1.11}
\end{equation}
which is linearly independent for $Re(s)> 0$, gives the unique Green
function which defines a bounded solution for $j$ of compact support. 
Note that for $Re(s)>0$ the solution $f_+$ is exponentially decaying
for large $x$ and $f_-$ is exponentially decaying for small $x$. For
small $x$ however, $f_+$ will be a linear combination $a(s)e^{-s x} +
b(s)e^{s x}$ which will in general grow exponentially. Similar
behavior is found for $f_-$.

Quasi-Normal mode frequencies $s_n$ can be defined as those complex 
numbers for which
\begin{equation}
  f_+(s_n,x)=c(s_n)f_-(s_n,x),
  \label{bs1.12}
\end{equation}
that is the two functions become linearly dependent, the Wronskian
vanishes and the Green function is singular! The corresponding
solutions $f_+(s_n,x)$ are called quasi eigenfunctions.

Are there such numbers $s_n$? From the boundedness of the solution in 
spacetime we know that the unique Green function must exist for
$Re(s)>0$. Hence $f_+,f_-$ are linearly independent for those values
of $s$. However, as solutions $f_+,f_-$ of the homogeneous
equation~(\ref{bs1.7b}) they have a unique continuation to the complex
$s$ plane. In~\cite{BB91} it is shown that for positive potentials
with compact support there is always a countable number of zeros of
the Wronskian with $Re(s)<0$.

What is the mathematical and physical significance of the quasi-normal 
frequencies $s_n$ and the corresponding quasi-normal functions $f_+$?
First of all we should note that because of $Re(s)<0$ the function
$f_+$ grows exponentially for small and large $x$! The corresponding
spacetime solution  $e^{s_n t} f_+(s_n, x) $ is therefore not a
physically relevant solution, unlike the normal modes. 

If one studies the inverse Laplace transformation and expresses $\chi$
as a complex line integral ($a>0$),
\begin{equation}
  \chi (t,x) = {1\over {2\pi i}}\int_{-\infty}^{+\infty}
  e^{(a+is)t}\hat{\chi}(a+is,x)ds,
  \label{bs1.13}
\end{equation}
one can deform the path of the complex integration and show that the
late time behavior of solutions can be approximated in finite parts of
the space by a finite sum of the form
\begin{equation}
  \chi(t,x)\sim\sum_{n=1}^N\ a_n e^{(\alpha_n+i\beta_n)t}
  f_+(s_n,x).
  \label{bs1.14}
\end{equation}
Here we assume that $Re(s_{n+1})<Re(s_n)<0$, $s_n=\alpha_n+i\beta_n$. 
The approximation $\sim$ means that if we choose $x_0$, $x_1$,
$\epsilon$ and $t_0$ then there exists a constant
$C(t_0,x_0,x_1,\epsilon)$ such that
\begin{equation}
  \left \vert \chi(t,x)-\sum_{n=1}^N a_n e^{(\alpha_n+i\beta_n)t}
    f_+(s_n,x) \right \vert \le Ce^{(-|\alpha_{N+1}|+\epsilon )t}
  \label{bs1.15}
\end{equation}
holds for $t>t_0$, $x_0<x<x_1$, $\epsilon>0$ with
$C(t_0,x_0,x_1,\epsilon)$ independent of $t$. The constants $a_n$
depend only on the data~\cite{BB91}! This implies in particular that
all solutions defined by data of compact support decay exponentially
in time on spatially bounded regions. The generic leading order decay
is determined by the quasi-normal mode frequency with the largest real
part $s_1$, i.e.\ slowest damping. On finite intervals and for late
times the solution is approximated by a finite sum of quasi
eigenfunctions~(\ref{bs1.14}).

It is presently unclear whether one can strengthen~(\ref{bs1.15}) to a
statement like~(\ref{bs1.2}), a pointwise expansion of the late time
solution in terms of quasi-normal modes. For one particular potential
(P\"oschl-Teller) this has been shown by Beyer~\cite{Horst}.
  
Let us now consider the case where the potential is positive for all
$x$, but decays near infinity as happens for example for the wave
equation on the static Schwarzschild spacetime. Data of compact
support determine again solutions which are bounded~\cite{KW87}.
Hence we can proceed as before. The first new point  concerns the
definitions of $f_\pm$. It can be shown that the homogeneous
equation~(\ref{bs1.7b}) has for each real positive $s$ a unique
solution $f_+(s,x)$ such that $\lim_{x\to\infty} (e^{sx}f_+(s,x))=1$
holds and correspondingly for $f_-$. These functions are uniquely
determined, define the correct Green function and have analytic
continuations onto the complex half plane $Re(s)>0$.

It is however quite complicated to get a good representation of these 
functions. If the point at infinity is not a regular singular point, we
do not even get converging series expansions for $f_\pm$. (This is
particularly serious for values of $s$ with negative real part because
we expect exponential growth in $x$).

The next new feature is that the analyticity properties of $f_\pm$ in
the complex $s$ plane depend on the decay of the potential. To obtain
information about analytic continuation, even use of analyticity
properties of the potential in $x$ is made! Branch cuts may occur. 
Nevertheless in a lot of cases an infinite number of quasi-normal mode 
frequencies exists.

The fact that the potential never vanishes may, however, destroy the 
exponential decay in time of the solutions and therefore the essential
properties of the quasi-normal modes. This probably happens if the
potential decays slower than exponentially. There is, however, the
following way out: Suppose you want to study a solution determined by
data of compact support from $t=0$ to some large finite time $t=T$. 
Up to this time the solution is -- because of domain of dependence 
properties -- completely independent of the potential for sufficiently 
large $x$. Hence we may see an exponential decay of the
form~(\ref{bs1.14}) in a time range $t_1<t<T$. This is the behavior
seen in numerical calculations. The situation is similar in the case
of $\alpha$-decay in quantum mechanics. A comparison of quasi-normal
modes of wave equations and resonances in quantum theory can be found
in the appendix, see section~\ref{appendix_1}.


\newpage

\section{Quasi-Normal Modes of Black Holes}
\label{section_3}

One of the most interesting aspects of gravitational wave detection 
will be the connection with the existence of black
holes~\cite{Thorne98}. Although there are presently several indirect
ways of identifying black holes in the universe, gravitational waves
emitted by an oscillating black hole will carry a unique fingerprint
which would lead to the direct identification of their existence.

As we mentioned earlier, gravitational radiation from black hole
oscillations exhibits certain characteristic frequencies which are independent of the
processes giving rise to these oscillations. These ``quasi-normal''
frequencies are directly connected to the parameters of the black hole
(mass, charge and angular momentum) and for stellar mass black holes
are expected to be inside the bandwidth of the constructed
gravitational wave detectors.

The perturbations of a Schwarzschild black hole reduce to a simple
wave equation which has been studied extensively. The wave equation
for the case of a Reissner-Nordstr\"om black hole is more or less
similar to the Schwarzschild case, but for Kerr one has to solve a
system of coupled wave equations (one for the radial part and one for
the angular part). For this reason the Kerr case has been studied less
thoroughly. Finally, in the case of Kerr-Newman black holes we face
the problem that the perturbations cannot be separated in their
angular and radial parts and thus apart from special
cases~\cite{KDK93} the problem has not been studied at all.


\subsection{Schwarzschild Black Holes}
\label{section_3_1}

The study of perturbations of Schwarzschild black holes assumes a
small perturbation $h_{\mu\nu}$ on a static spherically symmetric
background metric
\begin{equation}
  ds^2 =g_{\mu \nu}^0 dx^{\mu}dx^{\nu} = -e^{v(r)}dt^2 +
  e^{\lambda(r)}dr^2 + r^2 \left( d\theta^2 + \sin^2 \theta d\phi^2
  \right),
  \label{metric1}
\end{equation}
with the perturbed metric having the form
\begin{equation}
  g_{\mu\nu} = g_{\mu\nu}^0 + h_{\mu\nu},
  \label{scw_pert}
\end{equation}
which leads to a variation of the Einstein equations i.e.
\begin{equation}
  \delta G_{\mu\nu} = 4 \pi \delta T_{\mu\nu}.
\end{equation}
By assuming a decomposition into tensor spherical harmonics for each
$h_{\mu\nu}$ of the form
\begin{equation}
  \chi (t,r,\theta, \phi) = \sum_{\ell m}{\chi_{\ell m}(r, t)\over r}
  Y_{\ell m}(\theta, \phi),
  \label{harmonics}
\end{equation}
the perturbation problem is reduced to a single wave equation, for
the function $\chi_{\ell m}(r,t)$ (which is a combination of the
various components of $h_{\mu\nu}$). It should be pointed out that
equation~(\ref{harmonics}) is an expansion for scalar quantities only. From the 10 independent components of the $h_{\mu\nu}$ only
$h_{tt}$, $h_{tr}$, and $h_{rr}$ transform as scalars under rotations.
The $h_{t \theta}$, $h_{t\phi}$, $h_{r\theta}$, and $h_{r\phi}$
transform as components of two-vectors under rotations and can be
expanded in a series of vector spherical harmonics while the
components $h_{\theta\theta}$, $h_{\theta\phi}$, and $h_{\phi\phi}$
transform as components of a $2\times2$ tensor and can be expanded in
a series of tensor spherical harmonics (see~\cite{TC67, Zer70, Monc74c}
for details). There are two classes of vector spherical harmonics
({\bf polar} and {\bf axial}) which are build out of combinations of
the Levi-Civita volume form and the gradient operator acting on the
scalar spherical harmonics. The difference between the two families is
their parity. Under the parity operator $\pi$ a spherical harmonic
with index $\ell$ transforms as $(-1)^\ell$, the polar class of
perturbations transform under parity in the same way, as $(-1)^\ell$,
and  the axial perturbations as $(-1)^{\ell+1}$\footnote{In the
literature the {\em polar} perturbations are also called {\em
even-parity} because they are characterized by their behavior under
parity operations as discussed earlier, and in the same way the {\em
axial} perturbations are called {\em odd-parity}. We will stick to the
polar/axial terminology since there is a confusion with the
definition of the parity operation, the reason is that to most people, 
the words ``even'' and ``odd'' imply that a mode transforms under
$\pi$ as $(-1)^{2n}$ or $(-1)^{2n+1}$ respectively (for $n$ some
integer). However only the polar modes with even $\ell$ have even
parity and only axial modes with even $\ell$ have odd parity. If
$\ell$ is odd, then polar modes have odd parity and axial modes have
even parity. Another terminology is to call the polar perturbations
{\em spheroidal} and the axial ones {\em toroidal}. This definition is
coming from the study of stellar pulsations in Newtonian theory and
represents the type of fluid motions that each type of perturbation
induces. Since we are dealing both with stars and black holes we will
stick to
 the polar/axial terminology.}. Finally, since we are
dealing with spherically symmetric spacetimes the solution will be
independent of $m$, thus this subscript can be omitted.

The radial component of a perturbation outside the event horizon 
satisfies the following wave equation,
\begin{equation}
  {{\partial^2}\over {\partial t^2}} \chi_\ell + 
  \left( -{{\partial^2}\over {\partial r_*^2}} + V_\ell(r)
  \right)\chi_\ell = 0,
  \label{qnmwave}
\end{equation}
where $r_*$ is the ``tortoise'' radial coordinate defined by
\begin{equation}
  r_*=r+2M \log (r/2M-1),
\end{equation}
and $M$ is the mass of the black hole.

For ``axial'' perturbations
\begin{equation}
  V_\ell(r)=\left(1 - {2M\over r}\right) \left[ {{\ell(\ell+1)}\over
  r^2} + {{2\sigma M} \over r^3} \right]
  \label{rwpot}
\end{equation}
is the effective potential or (as it is known in the literature)
Regge-Wheeler potential~\cite{RW57}, which is a single potential
barrier with a peak around $r=3M$, which is the location of the
unstable photon orbit. The form~(\ref{rwpot}) is true even if we
consider scalar or electromagnetic test fields as perturbations.
The parameter $\sigma$ takes the values 1 for scalar perturbations, 0
for electromagnetic perturbations, and $-3$ for gravitational
perturbations and can be expressed as $\sigma=1-s^2$, where $s=0, 1,
2$ is the spin of the perturbing field.

For ``polar'' perturbations the effective potential was derived by
Zerilli~\cite{Zer70} and has the form
\begin{equation}
  V_\ell(r)=\left( 1 - \frac{2M}r \right) \frac{2n^2(n+1)r^3 + 6n^2Mr^2 +
  18nM^2r +18M^3}{r^3(nr+3M)^2},
\label{zerpot}
\end{equation}
where
\begin{equation}
  2 n= (\ell-1)(\ell+2).
\end{equation}
Chandrasekhar~\cite{Chandra75} has shown that one can transform the 
equation~(\ref{qnmwave}) for ``axial'' modes to the corresponding 
one for ``polar'' modes via a transformation involving differential
operations. It can also be shown that both forms are connected to the
Bardeen-Press~\cite{BP73} perturbation equation derived via the
Newman-Penrose formalism. The potential $V_\ell(r_*)$ decays
exponentially near the horizon, $r_*\to-\infty$, and as $r_*^{-2}$ for
$r_*\to +\infty$. 

From the form of equation~(\ref{qnmwave}) it is evident that the study 
of black hole perturbations will follow the footsteps of the theory
outlined in section~\ref{section_2}.

Kay and Wald~\cite{KW87} have shown that solutions with data of
compact support are bounded. Hence we know that the time independent
Green function $G(s,r_*,r_*')$ is analytic for $Re(s)>0$. The
essential difficulty is now to obtain the solutions $f_\pm$
(cf. equation~(\ref{bs1.9})) of the equation 
\begin{equation}
  s^2\hat{\chi} -\hat{\chi}''+ V \hat{\chi} = 0,
  \label{wave1}
\end{equation}
(prime denotes differentiation with respect to $r_*$)
which satisfy for real, positive $s$:
\begin{equation}
  f_+\sim e^{-s r_*} \quad \mbox{for }  r_*\to\infty,
  \quad \quad f_-\sim e^{+r_* x} \quad \mbox{for } r_*\to -\infty. 
  \label{bconditions}
\end{equation}
To determine the quasi-normal modes we need the analytic continuations
of these functions.

As the horizon ($r_*\to\-\infty$) is a regular singular point of  
(\ref{wave1}), a representation of $f_-(r_*,s)$ as a converging series
exists. For $M={1\over 2}$ it reads:
%
%
\begin{equation}
  f_-(r,s)=(r-1)^s\sum_{n=0}^\infty a_n(s)(r-1)^n.
  \label{exp_hor}
\end{equation}
The series converges for all complex $s$ and $|r-1|<1$~\cite{Persides}.
(The analytic extension of $f_-$ is investigated in~\cite{JC85}.)
The result is that $f_-$ has an extension to the complex $s$ plane
with  poles only at negative real integers. The representation of
$f_+$ is more complicated: Because infinity is a singular point no
power series expansion like~(\ref{exp_hor}) exists. A representation
coming from the iteration of the defining integral equation is given
by Jensen and Candelas~\cite{JC85}, see also~\cite{NS92}. It turns out
that the continuation of $f_+$ has a branch cut $Re(s)\leq 0$ due to
the decay $r^{-2}$ for large $r$~\cite{JC85}.
                                                      
The most extensive mathematical investigation of quasi-normal modes of
the Schwarzschild solution is contained in the paper by Bachelot and
Motet-Bachelot~\cite{BB91}. Here the existence of an infinite number
of quasi-normal modes is demonstrated. Truncating the
potential~(\ref{rwpot}) to make it of compact support leads to the
estimate~(\ref{bs1.15}).

The decay of solutions in time is not exponential because of the weak
decay of the potential for large $r$. At late times, the quasi-normal
oscillations are swamped by the radiative tail~\cite{Price72a, Price72b}.
This tail radiation is of interest in its own right since it originates 
on the background spacetime. The first authoritative study of nearly 
spherical collapse, exhibiting radiative tails, was performed by
Price~\cite{Price72a, Price72b}. 

Studying the behavior of a massless scalar field propagating on a
fixed Schwarzschild background, he showed that the field dies off with
the power-law tail,
\begin{equation}
  \chi(r,t) \sim t^{-(2\ell+P+1)},
\end{equation}
at late times, where $P=1$ if the field is initially static, and $P=2$
otherwise. This behavior has been seen in various calculations, for
example the gravitational collapse simulations by Cunningham, Price
and Moncrief~\cite{CPMa, CPMb, CPMc}. Today it is apparent in any
simulation involving evolutions of various fields on a black hole
background including Schwarzschild,
Reissner-Nordstr\"om~\cite{GPP94a}, and Kerr~\cite{KLP96, KLPA97}. 
It has also been observed in simulations of axial oscillations of
neutron stars~\cite{ak96}, and should also be present for polar
oscillations. Leaver~\cite{Leaver86} has studied in detail these tails
and associated this power low tail with  the branch-cut integral
along the negative imaginary $\omega$ axis in the complex
$\omega$ plane. His suggestion that there will be radiative tails
observable at ${\cal J}^+$ and ${\cal H}^+$ has been verified by 
Gundlach, Price, and Pullin~\cite{GPP94a}. Similar results were
arrived at recently by Ching et al.~\cite{suen2} in a more extensive
study of the late time behavior. In a nonlinear study Gundlach, Price,
and Pullin~\cite{GPP94b} have shown that tails develop even when the
collapsing field fails to produce a black hole.
Finally, for a study of tails in the presence of a cosmological constant
refer to~\cite{BCKL97}, while for a recent study, using analytic methods,
of the late-time tails of linear scalar fields outside Schwarzschild and
Kerr black holes refer to~\cite{Barak99, BO99}.

Using the properties of the waves at the horizon and infinity 
given in equation~(\ref{bconditions}) one can search for the
quasi-normal mode frequencies since practically the whole problem has
been reduced to a boundary value problem with $s=i\omega$ being the
complex eigenvalue. The procedure and techniques used to solve the
problem will be discussed later in section~\ref{section_6}, but it is
worth mentioning here a simple approach to calculate the QNM
frequencies proposed by Schutz and Will~\cite{SW85}. The approach is
based on the standard WKB treatment of wave scattering on the peak of
the potential barrier, and it can be easily shown that the complex
frequency can be estimated from the relation
\begin{equation}
  (M\omega_n)^2 = V_\ell(r_0) - i \left(n+{1\over 2}\right) \left[-2
    {{d^2V_\ell(r_0)} \over {dr_*^2}}\right]^{1/2},
  \label{qnmsw}
\end{equation} 
where $r_0$ is the peak of the potential barrier. For $\ell=2$ and
$n=0$ (the fundamental mode) the complex frequency is $M\omega \approx
(0.37, -0.09)$, which for a $10 M_\odot$ black hole corresponds to a
frequency of $1.2$~kHz and damping time of $0.55$~ms. A few more QNM
frequencies for $\ell = 2, 3$ and 4 are listed in table~\ref{table1}.

\begin{table}[hptb]
  \begin{center}
    \begin{tabular}{|l|ll|ll|ll|}
      \hline
      n & $\ell=2$ & & $\ell=3$ & & $\ell=4$ & \cr
      \hline
      0 &  0.37367 & -0.08896 i & 0.59944 & -0.09270 i & 0.80918 &
      -0.09416 i \\
      1 &  0.34671 & -0.27391 i & 0.58264 & -0.28130 i & 0.79663 &
      -0.28443 i \\
      2 &  0.30105 & -0.47828 i & 0.55168 & -0.47909 i & 0.77271 &
      -0.47991 i \\
      3 &  0.25150 & -0.70514 i & 0.51196 & -0.69034 i & 0.73984 &
      -0.68392 i \\
      \hline
    \end{tabular}
 
\end{center}
  \caption{\it The first four QNM frequencies ($\omega M$) of the
    Schwarzschild black hole for $\ell =2, 3$, and
    $4$\protect~\protect\cite{Leaver85}. The frequencies are given in
    geometrical units and for conversion into kHz one should multiply
    by $2\pi (5142Hz)\times (M_\odot / M)$.}
  \label{table1}
\end{table}

\begin{figure}[hptb]   
  \epsfysize=6.5cm  
  \epsfxsize=8.5cm 
  \centerline{\epsffile{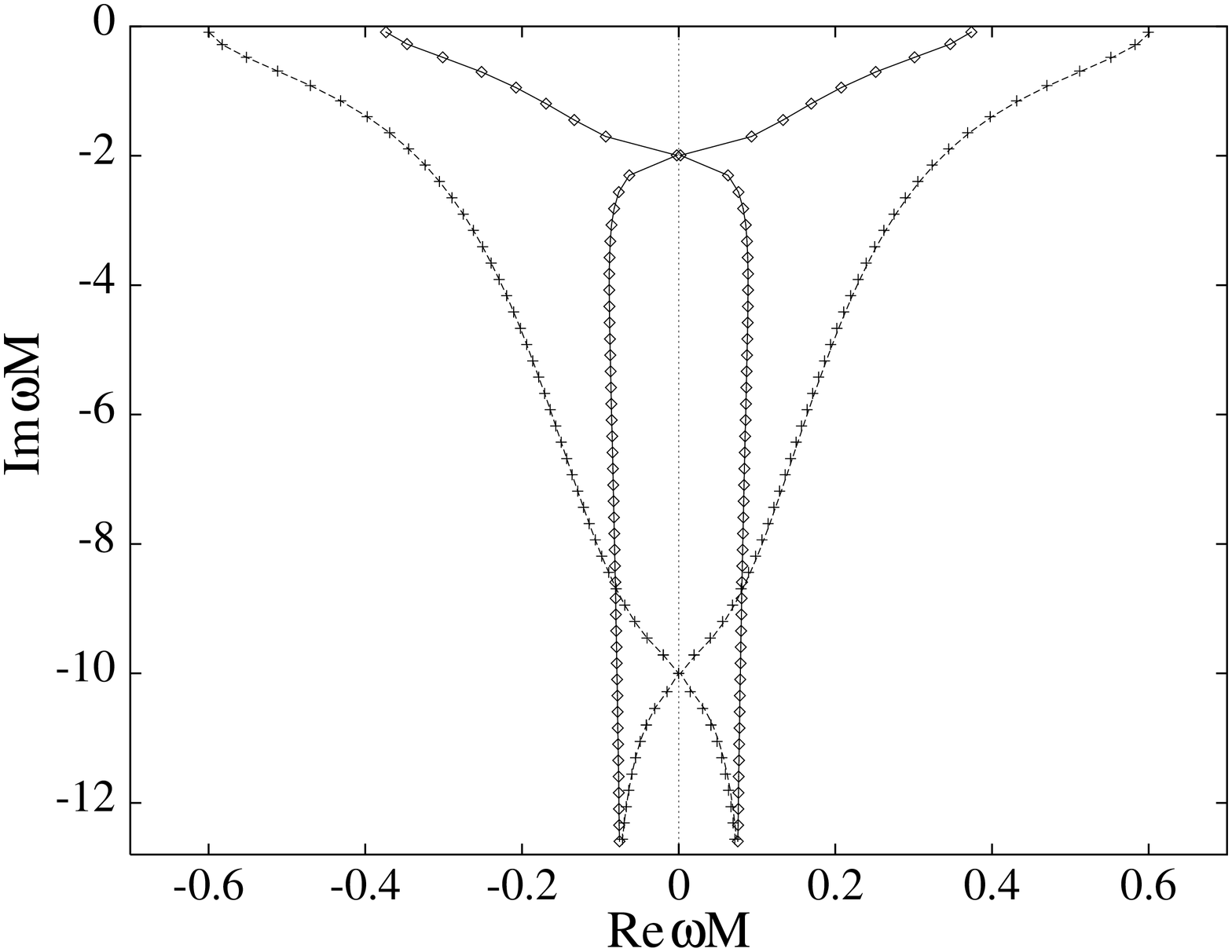}}  
     \caption{\it The spectrum of QNM for a Schwarzschild black-hole, for
     $\ell=2$ (diamonds) and $\ell=3$ (crosses)~\cite{AL92}. The 9th mode 
     for $\ell=2$ and the 41st for $\ell=3$ are ``special'', i.e.\ the
     real part of the frequency is zero ($s=i\omega$).}
  \label{bhmodes}   
\end{figure}

Figure~\ref{bhmodes} shows some of the modes of the Schwarzschild
black hole. The number of modes for each harmonic index $\ell$ is
infinite, as was mathematically proven by Bachelot and
Motet-Bachelot~\cite{BB91}. This was also implied in an earlier work
by Ferrari and Mashhoon~\cite{FM84b}, and it has been seen in the
numerical calculations in~\cite{AL92, Nollert93}. It can be also seen
that the imaginary part of the frequency grows very quickly. This
means that the higher modes do not contribute significantly in the
emitted gravitational wave signal, and this is also true for the
higher $\ell$ modes (octapole etc.).
As is apparent in figure~\ref{bhmodes} that there is a special 
purely imaginary QNM frequency. The existence of ``algebraically
special'' solutions for perturbations of Schwarzschild,
Reissner-Nordstr\"om and Kerr black holes were first pointed out by
Chandrasekhar~\cite{Chandra84}. It is still questionable whether these
frequencies should be considered as QNMs~\cite{Leaver90} and there is
a suggestion that the potential might become transparent for these
frequencies~\cite{Nils94}. For a more detailed discussion refer
to~\cite{LM96}.

As a final comment we should mention that as the order of the modes 
increases the real part of the frequency remains constant, while the 
imaginary part increases proportionally to the order of the mode. 
Nollert~\cite{Nollert93} derived the following approximate formula for
the asymptotic behavior of QNMs of a Schwarzschild black hole,
\begin{equation}
  M \omega_n \approx 0.0437 + {{\gamma_1}\over {(2n+1)^{1/2}}} + \dots -
  i \left[ -{1\over 8} (2n+1) + {{\gamma_1}\over {(2n+1)^{1/2}}} + \dots
  \right],
\label{asymqnm}
\end{equation}
where $\gamma_1 = 0.343$, $0.7545$ and $2.81$ for $\ell=$ 2, 3 and 6,
correspondingly, and $n \rightarrow \infty$. The above relation was
later verified in~\cite{nils93} and~\cite{Liu95}.

For large values of $\ell$ the distribution of QNMs is given
by~\cite{press71, FM84a, FM84b, Iyer87}
\begin{equation}
  3 \sqrt{3} M \omega_n \approx \ell+ {1\over2} - i \left( n+ {1\over
  2}\right).
\end{equation} 
For a mathematical proof refer to~\cite{BZ97}.

The perturbations of Reissner-Nordstr\"om black holes, due to the
spherical symmetry of the solution, follow the footsteps of the
analysis that we have presented in this section. Most of the work was
done during the seventies by
Zerilli~\cite{Zer74}, Moncrief~\cite{Monc74a, Monc74b} and later by
Chandrasekhar and Xanthopoulos~\cite{Chandra80, Xanth81}. For an
extensive discussion refer to~\cite{Chandra83}. We have again wave
equations of the form~(\ref{qnmwave}), one for each parity with
potentials which are like~(\ref{rwpot}) and~(\ref{zerpot}) plus extra
terms which relate to the charge of the black hole. An interesting
feature of the charged black holes is that any perturbation of the
gravitational (electromagnetic) field will also induce electromagnetic
(gravitational) perturbations. In other words, any perturbation of the
Reissner-Nordstr\"om spacetime will produce both electromagnetic and
gravitational radiation. Again it has been shown that the solutions
for the odd parity oscillations can be deduced from the solutions for
even parity oscillations and vice versa~\cite{Chandra80}. The QNM
frequencies of the  Reissner-Nordstr\"om black hole have been
calculated by Gunter~\cite{Gunter80}, Kokkotas, and
Schutz~\cite{KS88}, Leaver~\cite{Leaver90}, Andersson~\cite{nils93b},
and lately for the nearly extreme case by Andersson and
Onozawa~\cite{AO96}.


\subsection{Kerr Black Holes}
\label{section_3_2}

The Kerr metric represents an axisymmetric, black hole solution to the
source free Einstein equations. The metric in $(t,r,\theta,\varphi)$
coordinates is 
%
\begin{eqnarray}
  ds^2&=&-\left(1-{2Mr\over \Sigma}\right)dt^2 -{4Mar\sin^2\theta\over
    \Sigma} dtd\varphi +{\Sigma\over \Delta} dr^2  + \Sigma d\theta^2
  \nonumber \\
  &&+ \left(r^2+a^2 +{2Ma^2r\sin^2\theta\over \Sigma}\right)\sin^2\theta
  d\varphi^2,
  \label{kerr}
\end{eqnarray}
with
\begin{equation}
  \Delta=r^2-2Mr+a^2, \quad \quad \Sigma=r^2+a^2\cos^2\theta.
\end{equation}
$M$ is the mass and $0\leq a\leq M$ the rotational parameter of the
Kerr metric. The zeros of $\Delta$ are
\begin{equation}
  r_\pm=M\pm (M^2-a^2)^{1/2},
\end{equation}
and determine the horizons. For $r_+\leq r<\infty$ the spacetime
admits locally a timelike Killing vector. In the ergosphere region
\begin{equation}
  r_+\leq r<M+(m^2-a^2\cos^2\theta)^{1/2},
\end{equation}
the Killing vector $\partial / \partial t$ which is timelike at
infinity, becomes spacelike. The scalar wave equation for the Kerr
metric is 
%
\begin{eqnarray}
  & &\left[ {(r^2+a^2)^2\over\Delta} - a^2\sin^2\theta \right]
  {\partial^2\chi\over\partial t^2}+{4Mar\over\Delta}
  {\partial^2\chi\over\partial t\partial\varphi}+\left[{a^2\over\Delta}
  - {1\over\sin^2\theta}\right] {\partial^2\chi\over\partial \varphi^2}
  \nonumber \\
  &&- \Delta^{-\sigma} {\partial \over {\partial r}}
  \left(\Delta^{\sigma+1} {\partial \chi \over \partial r} \right)
  -{1\over\sin\theta}{\partial\over\partial\theta}
  \left(\sin\theta{\partial \chi \over\partial\theta}\right)
  - 2 \sigma \left[ {{a(r-M)}\over \Delta} + {i\cos\theta \over
  \sin^2\theta} \right]{\partial \chi \over \partial \varphi} \nonumber \\
  &&- 2 \sigma \left[ {M(r^2-a^2)\over \Delta} -r - i a \cos \theta\right]
  {\partial \chi \over \partial t} + \left( \sigma^2 \cot^2\theta -
  \sigma \right)\chi=0,
  \label{teu1}
\end{eqnarray}
where
$\sigma = 0, \pm 1, \pm 2$ for scalar, electromagnetic or gravitational 
perturbations, respectively. As the Kerr metric outside the  horizon
$(r>r_+)$ is globally hyperbolic, the Cauchy problem for the scalar
wave equation~(\ref{teu1}) is well posed for data on any Cauchy surface.
However, the coefficient of $\partial^2\chi / \partial\varphi^2$
becomes negative in the ergosphere. This implies that the time
independent equation we obtain after the Fourier or Laplace
transformation is not elliptic!

For linear hyperbolic equations with time independent coefficients, we
know that solutions determined by  data with compact support are
bounded by $ce^{\gamma t}$, where $\gamma$ is independent of the
data. It is not known whether all such solutions are bounded in time,
i.e.\ whether they are stable.

Assuming harmonic time behavior $\chi=e^{i\omega
t}\hat{\chi}(r,\theta,\varphi)$, a separation in angular and radial
variables was found by Teukolsky~\cite{Teu72}:
\begin{equation}
  \hat{\chi}(r,\theta,\varphi)=R(r,\omega) S(\theta,\omega)e^{im\varphi}.
\end{equation}
Note that in contrast to the case of spherical harmonics, the
separation is $\omega$-dependent. To be a solution of the wave
equation~(\ref{teu1}), the functions $R$ and $S$ must satisfy 
\begin{equation}
  {1\over \sin\theta}{d \over d \theta}\left[\sin\theta {d S\over d
  \theta} \right] \!+\! \left[a^2\omega^2\cos^2\theta + 2a\omega \sigma
  \cos\theta - {m^2 \!\! + \! \sigma^2 \!\! + \! 2m \sigma
  \cos\theta \over \sin^2\theta} \! + \! E \right] S = 0,
  \label{teu1a}
\end{equation}
\begin{equation}
  \Delta^{-\sigma}{d\over {d r}}\left[\Delta^{\sigma+1}{d R\over {d
  r}}\right]+ {1\over\Delta}\left[K^2 + 2 i \sigma(r-1)K -\Delta(4 i
  \sigma r\omega+\lambda)\right] R = 0,
  \label{teu1b}
\end{equation}
where $K = (r^2+a^2)\omega + am$, $\lambda = E + a^2\omega^2 +
2am\omega - \sigma(\sigma+1)$, and $E$ is the separation constant.

For each complex $a^2\omega^2$ and positive integer $m$,
equation~(\ref{teu1a}) together with the boundary conditions of
regularity at the axis, determines a singular Sturm-Liouville
eigenvalue problem. It has solutions for eigenvalues
$E(\ell,m^2,a^2\omega^2)$, $|m| \leq \ell$. The eigenfunctions are the
spheroidal (oblate) harmonics $S_{\ell|m|}(\theta)$. They exist for
all complex $\omega^2$. For real $\omega^2$ the spheroidal harmonics
are complete in the sense that any function  of $z=\cos\theta$,
absolutely integrable over the interval $[-1, 1]$, can be expanded
into spheroidal harmonics of fixed $m$~\cite{Seidel89}. Furthermore,
functions $A(\theta,\varphi)$ absolutely integrable over the sphere
can be expanded into 
\begin{equation} 
  A(\theta,\varphi)=\sum_{\ell=0}^\infty\sum_{m=-\ell}^{+\ell}
  A(\ell,m,\omega)S_{\ell|m|}(\theta)e^{im\varphi}.
\end{equation}
For general complex $\omega^2$ such an expansion is not possible. 
There is a countable number of ``exceptional values'' $\omega^2$  
where no such expansion exists~\cite{MS54}.

Let us pick one such solution $S_{\ell|m|}(\theta,\omega)$ and
consider some solutions $R(r,\omega)$ of~(\ref{teu1b}) with the
corresponding $E(\ell,|m|,\omega)$. Then $R_{\ell|m|}(r,\omega)
S_{\ell|m|}e^{im\varphi}e^{-i\omega t}$ is a solution
of~(\ref{teu1}). Is it possible to obtain ``all'' solutions by summing
over $\ell,m$ and integrating over $\omega$? For a solution in
spacetime for which a Fourier transform in time exists at any space
point (square integrable in time), we can expand the Fourier transform
in spheroidal harmonics because $\omega$ is real. The coefficient
$R(r,\omega,m)$ will solve equation~(\ref{teu1b}). Unfortunately, we
only know that a solution determined by data of compact support is
exponentially bounded. Hence we can only perform a Laplace
transformation.

We proceed therefore as in section~\ref{section_2}. Let
$\hat{\chi}(s,r,\theta,\varphi)$ be the Laplace transform of a
solution determined by data $\chi(t,r,\theta,\varphi)=0$ and
$\partial_t\chi(t,r,\theta,\varphi)=\rho$, while $\chi$ is analytic in
$s$ for real $s>\gamma\geq 0$, and has an analytic continuation onto
the half-plane $Re(s)\geq\gamma$. For real $s$ we can expand
$\hat{\chi}$ into a converging sum of spheroidal harmonics~\cite{MS54}
\begin{equation}
  \hat{\chi}(s,r,\theta,\varphi) = \sum_{\ell=0}^\infty
  \sum_{m=-\ell}^{+\ell} R(\ell,m,s)
  S_{\ell|m|}(-s^2,\theta)e^{im\varphi}. 
  \label{expan1}
\end{equation}
$R(\ell,m,s)$ satisfies the radial equation~(\ref{bs1.8})
with $i\omega=s$. This representation of $\hat{\chi}$ does, however,
not hold for all complex values in the half-plane on which it is
defined. Nevertheless is it true that for all values of $s$ which are
not exceptional an expansion of the form~(\ref{expan1})
exists\footnote{The definition of $S_{\ell|m|}(-s^2,\theta)$ on the
  complex plane is made unique by fixing certain conventions about the
  branch cuts. The exceptional points are the beginnings of branch
  cuts. $R(l,m,s)$ as a solution of~(\ref{teu1b}) is defined also at
  the exceptional points; just the series does not exist there.}.

To define quasi-normal modes we first have to define the correct Green 
function of~(\ref{teu1b}) which determines $R(\ell,m,s)$ from the data
$\rho$. As usual, this is done by prescribing decay for real $s$ for
two linearly independent solutions $^\pm R(\ell,m,s)$ and analytic
continuation. Out of the Green function for $R(\ell,m,s)$ and
$S_{\ell|m|}(-s^2,\theta)$ for real $s$ we can build the Green
function $G(s,r,r',\theta,\theta',\phi,\phi')$ by a series
representation like~(\ref{expan1}). Analytic continuation defines $G$
on the half plane $Re(s)>\gamma$. For non exceptional $s$ we have a
series representation. (We must define $G$ by this complicated
procedure because the partial differential operator corresponding
to~(\ref{teu1a}),~(\ref{teu1b}) is not elliptic in the ergosphere.)

Normal and quasi-normal modes appear as poles of the analytic
continuation of $G$. Normal modes are determined by poles with
$Re(s)>0$ and quasi-normal modes by $Re(s)<0$. Suppose all such values
are different from the exceptional values. Then we have always the
series expansion of the Green function near the poles and we see that
they appear as poles of the radial Green function of $R(\ell,m,s)$. 

To relate the modes to the asymptotic behavior in time we study the
inverse Laplace transform and deform the integration path to include
the contributions of the poles. The decay in time is dominated either
by the normal mode with the largest (real) eigenfrequency or the
quasi-normal mode with the largest negative real part. 

It is apparent that the calculation of the QNM frequencies of the
Kerr black hole is more involved than the Schwarzschild and
Reissner-Nordstr\"om cases. This is the reason that there have only
been a few attempts~\cite{Leaver85, SI90, Kokkotas91,Onoz97} in this
direction.

The quasi-normal mode frequencies of the Kerr-Newman black hole have
not yet been calculated, although they are more general than all other
types of perturbation. The reason is the complexity of the
perturbation equations and, in particular, their non-separability. This
can be understood through the following analysis of the perturbation
procedure. The equations governing a perturbing massless field of spin
$\sigma$ can be written as a set of $2\sigma+1$ wavelike equations in
which the various different helicity components of the perturbing
field are coupled not only with each other but also with the curvature
of the background space, all with four independent variables as
coordinates over the manifold. The standard problem is to decouple the
$2\sigma+1$ equations or at least some physically important subset of
them and then to separate the decoupled equations so as to obtain
ordinary differential equations which can be handled by one of the
previously stated methods. For a discussion and estimation of the QNM
frequencies in a restrictive case refer to~\cite{KDK93}.


\subsection{Stability and Completeness of Quasi-Normal Modes}
\label{section_3_3}

From the normal modes one can learn a lot about stability. Take as an
example linear stellar oscillation within the framework of Newton's
theory of gravity. As outlined  at the beginning of
section~\ref{section_2}, we have a sequence of normal modes $\omega_n$
with $\omega^2$ real. The general solution is a convergent linear
combination of the corresponding eigenfunctions. Hence all
eigenfunctions are bounded if  $\omega$ is purely imaginary,
i.e.\ $\omega^2<0$.

Therefore the spectrum contains all the information about
stability. To discuss stability for systems with quasi-normal modes,
let us consider a case like equation~(\ref{bs1.5}) with the assumption
that $V$ is of compact support but not necessarily positive. 

Data of compact support define solutions which grow at most
exponentially in time
\begin{equation}
  |\phi(t,x)|< c e^{at},
\label{stabil}
\end{equation}
where $a$ is independent of the data. As outlined in
appendix~\ref{appendix_1}, eigenvalues necessarily have $s_n>0$ and
the eigenfunctions determine solutions growing exponentially in time. If
no eigenvalues exist, the solution can not grow
exponentially. Polynomial growth is still possible and related to the
properties of the Laplace transform of the Green function at $s=0$. As 
the potential has compact support, the functions $f_\pm(s,x)$ are
analytic for all $s$. Hence, the Green function can at most have a
pole at $s=0$. A pole of order two and higher implies polynomial
growth in time. If the potential is positive, energy conservation
shows that the field can grow at most linearly in time and therefore
we can have at most a pole of order 2 at $s=0$.  

If we define stability as boundedness in time for all solutions
with data of compact support, properties of quasi-normal modes can not
decide the stability issue. However, the appearance of a normal mode
proves instability. If the support of the potential is not compact
everything becomes more complicated. In particular, it is a non
trivial problem to obtain the behavior of the Green function at $s=0$.

In the case of the Schwarzschild black hole, stability is demonstrated 
by Kay and Wald~\cite{KW87} who showed the boundedness of all solutions 
with data of compact support.

The issue is more subtle for Kerr. There is a conserved energy, but
because of the ergoregion its integrand is not positive definite,
hence the conserved energy could be finite while the field still might
grow exponentially in parts of the spacetime. Papers by Press and
Teukolsky~\cite{PT73}, Hartle and Wilkins~\cite{HW74}, and
Stewart~\cite{Stewart} try to exclude the existence of an
exponentially growing normal mode. Their work makes the stability very
plausible but is not as conclusive as the Wald-Kay result. However
this is a delicate issue as we see if, for example, we multiply the
Regge-Wheeler potential by a factor $\epsilon$: For any $\epsilon > 0$
we obtain an infinite number of QNMs, for $\epsilon=0$, however there
is no QNM! Whiting~\cite{Whiting} has proven that there are no
exponentially growing modes, and in his proof he showed that the
growth of the  modes is at most linear. Recent numerical evolution
calculations~\cite{KLP96, KLPA97} for slowly and fast rotating Kerr
black holes pick up all the expected features (QNM ringing, tails) and
show no sign of exponential growth. It should be noted that the
massive scalar perturbations of Kerr are known to be
unstable~\cite{DDR76, ZE79, Det80}. These unstable modes are known to be
very slowly growing (with growth times similar to the age of universe).

Let us finally turn to the ``completeness of QNMs''. A general
mathematical theorem (spectral theorem) implies that for systems like
strings or membranes the general solution can be expanded into a
converging sum of normal modes. A similar result can not be expected
for QNMs, the reasons are given  in section~\ref{section_2}. There is,
however, the possibility that an infinite sum of the
form~(\ref{bs1.14}) will be a representation of a solution for late
times. This property has been shown by Beyer~\cite{Horst} for the
P\"oschl-Teller potential which has a similar form as the potential
on Schwarzschild~(\ref{rwpot}). The main difference is its exponential
decay at both ends. In~\cite{NP99} Nollert and Price propose a
definition of completeness and show its adequateness for a particular
model problem. There are also systematic studies~\cite{suen1} about
the relation between the structure of the QNM's of the Klein-Gordon
equation and the form of the potential. In these studies there is a
discussion on both the requirements for QNMs to form a
complete set and the definition of completeness.


\newpage

\section{Quasi-Normal Modes of Relativistic Stars}
\label{section_4}

Pulsating stars are important sources of information for
astrophysics. Nearly every star undergoes some kind of pulsation
during its evolution from the early stages of formation until the very
late stages, usually the catastrophic creation of a compact object
(white dwarf, neutron star or black hole). Pulsations of supercompact
objects are of great importance for relativistic astrophysics since
these pulsations are accompanied by the emission of gravitational
radiation. Neutron star oscillations were also proposed to explain the
quasi-periodic variability found in radio-pulsar and X-ray burster
signals~\cite{vanhorn, MVH88}. In this chapter we shall discuss
various features of neutron star non-radial pulsations i.e.\ the
various modes of pulsation,  mode excitation, detection probability
and the possibility to  extract information (to estimate, for example,
the radius, mass and stellar equation of state) from the detection of
the associated gravitational waves. It is not in our plans to discuss
rotating relativistic stars; the interested reader should refer to
another review in this journal~\cite{Nick98}. Radial oscillations are
also not discussed since they are not interesting for gravitational
wave research.


\subsection{Stellar Pulsations: The Theoretical Minimum}
\label{section_4_1}

For the study of stellar oscillations we shall consider a spherically
symmetric and static spacetime which can be described by the
Schwarzschild solution outside the star, see
equation~(\ref{metric1}). Inside the star, assuming that the stellar
material is behaving like an ideal fluid, we define the energy momentum
tensor
\begin{equation}
  T_{\mu\nu}=(\rho +p)u_\mu u_\nu + p g_{\mu \nu},
\end{equation}
where $p(r)$ is the pressure, $\rho(r)$ is the total energy
density. Then from the conservation of the energy-momentum and the
condition for hydrostatic equilibrium we can derive the
Tolman-Oppenheimer-Volkov (TOV) equations for the interior of a
spherically symmetric star in equilibrium. Specifically,
\begin{equation}
  e^{-\lambda} = 1 - {2 m(r) \over r},
\end{equation}
and the ``mass inside radius $r$'' is represented by
\begin{equation}
  m(r) = 4\pi \int_0^r \rho r^2 dr.
\end{equation}
This means that the total mass of the star is $M=m(R)$, with $R$ being
the star's radius. To determine a stellar model we must solve
\begin{equation}
  {dp \over dr} = - {\rho + p \over 2} {d\nu \over dr},
\end{equation}
where
\begin{equation}
  {d\nu \over dr} = {2 e^\lambda(m + 4\pi p r^3) \over r^2}.
\end{equation}
These equations should of course, be supplemented with  an equation of
state $p = p (\rho, \dots)$ as input. Usually is sufficient to use a
one-parameter equation of state to model neutron stars, since the
typical thermal energies are much smaller than the Fermi energy. The
polytropic equation of state $p=K \rho^{1+1/N}$ where $K$ is the
polytropic constant and $N$ the polytropic exponent, is used in most
of the studies. The existence of a unique global solution of the
Einstein equations for a given equation of state and a given value of
the central density has been proven by Rendall and
Schmidt~\cite{RS91}.

If we assume a small variation in the fluid or/and in the spacetime
we must deal with the perturbed Einstein equations 
\begin{equation}
  \delta \left( G^{\mu}_{\nu} -  {{8 \pi G}\over {c^4}}
    T^{\mu}_{\nu}\right) = 0,
\end{equation}
and the variation of the fluid equations of motion 
\begin{equation}
  \delta \left( T^{\mu}_{\nu ; \mu} \right) = 0,
\end{equation}
while the perturbed metric will be given by equation~(\ref{scw_pert}).

Following the procedure of the previous section one can decompose the
perturbation equations into spherical harmonics. This decomposition
leads to two classes of oscillations according to the parity of the
harmonics (exactly as for the black hole case). The first ones called
{\em even} (or spheroidal, or polar) produce spheroidal deformations
on the fluid, while the second are the {\em odd} (or toroidal, or
axial) which produce toroidal deformations. 

For the {\em  polar} case one can use certain combinations of the
metric perturbations as unknowns, and the linearized field equations
inside the star will be equivalent to the following system  of three
wave equations for unknowns $S, F, H$:
\begin{equation}
  -{1\over c^2}{\partial^2S\over \partial^2t}
  +{\partial^2S\over \partial^2r_*} + L_1(S,F,\ell) = 0,
  \label{starper1}
\end{equation}
\begin{equation}
  -{1\over c^2}{\partial^2F\over \partial^2t}
  +{\partial^2F\over \partial^2r_*} + L_2(S,F,H,\ell) = 0,
  \label{starper2}
\end{equation}
\begin{equation}
  -{1\over (c_s)^2}{\partial^2H\over \partial^2t}
  +{\partial^2H\over \partial^2r_*} + L_3(H,H',S,S',F,F',\ell) = 0,
  \label{starper3}
\end{equation}
and the constraint
\begin{equation}
  {\partial^2F\over \partial^2r_*}+L_4(F,F',S,S',H,\ell) = 0.
  \label{starconstr}
\end{equation}
%
The linear functions $L_i$, ($i=1, 2, 3, 4$) depend on the background
model and their explicit form can be found in~\cite{KES93, aaks}. The
functions $S$ and $F$ correspond to the perturbations of the spacetime
while the  function $H$ is proportional to the density perturbation
and is only defined on the background star. With $c_s$ we define the
speed of sound and with a prime we denote differentiation  with
respect to $r_*$:
\begin{equation}
  {\partial \over {\partial r_*}} = e^{(v-\lambda)/2}
  {\partial \over {\partial r}}.
\end{equation}
Outside the star there are only perturbations of the spacetime. These
are described by a single wave equation, the Zerilli equation
mentioned in the previous section, see equations~(\ref{qnmwave})
and~(\ref{zerpot}). In~\cite{KES93} it was shown that (for background
stars whose boundary density is positive) the above system -- together
with the geometrical transition conditions at the boundary of the star
and regularity conditions at the center -- admits a well posed Cauchy
problem. The constraint is preserved under the evolution. We see that
two variables propagate along light characteristics and the density
$H$ propagates with the sound velocity of the background star.

It is possible to eliminate the constraint -- first done by
Moncrief~\cite{Monc74c} -- if one solves the
constraint~(\ref{starconstr}) for $H$ and puts the corresponding
expression into $L_2$. (The characteristics for $F$ change then to
sound characteristics inside the star and light characteristics
outside.) This way one has just to solve two coupled wave equations
for $S$ and $F$ with unconstrained data, and to calculate $H$ using the
constraint from the solution of the two wave equations. Again the
explicit form of the equation can be found in~\cite{aaks}.

Turning next to quasi-normal modes in the spirit of
section~\ref{section_2}, we can Laplace transform  the two wave
equations and obtain a system of ordinary differential equations which 
is of fourth order. The Green function can be constructed from
solutions of the homogeneous equations (having the appropriate
behavior at the center and infinity)  and its analytic continuation
may have poles  defining the quasi-normal mode frequencies.   

From the form of the above equations one can easily see two limiting
cases. Let us first assume that the gravitational field is very
weak. Then equation~(\ref{starper1}) and~(\ref{starper2}) can be
omitted (actually $S \to 0$ in the weak field
limit~\cite{Thorne69b, aaks}) and we find that one equation is enough
to describe (with acceptable accuracy) the oscillations of the
fluid. This approach is known as the Cowling
approximation~\cite{cowl41}. Inversely, we can assume that the
coupling between the two equations~(\ref{starper1})
and~(\ref{starper2}) describing the spacetime perturbations with the
equation~(\ref{starper3}) is weak and consequently derive  all the
features of the spacetime perturbations from only the two of
them. This is what is called the ``inverse Cowling approximation''
(ICA)~\cite{AKS96}.

For the {\em axial} case the perturbations reduce to a single wave
equation for the spacetime perturbations which describes toroidal
deformations
\begin{equation}
  -{1\over c^2}{\partial^2 X\over \partial^2t}
  +{\partial^2X\over \partial^2r_*} 
  + {e^v\over r^3}\left[\ell(\ell+1)r+r^3(\rho-p)-6M\right] = 0,
  \label{starper4}
\end{equation}
where $X\sim h_{r\phi}$. Outside the star, pressure and density are
zero and this equation is reduced to the Regge-Wheeler equation, see
equations~(\ref{qnmwave}) and~(\ref{zerpot}). In Newtonian theory, if
the star is non-rotating and the static model is a perfect fluid
(i.e.\ shear stresses are absent), the {\em axial} oscillations are a
trivial solution of zero frequency to the perturbation equations and
the variations of pressure and density are zero. Nevertheless, the
variation of the velocity field is not zero and produces
non-oscillatory eddy motions. This means that there are no oscillatory
velocity fields. In the relativistic case the picture is
identical~\cite{TC67} nevertheless; in this case there are still QNMs,
the ones that we will describe later as ``{\em spacetime or
$w$-modes}''~\cite{kokkotas94}.

When the star is set in slow rotation then the axial modes are no longer
degenerate, but instead a new family of modes emerges, the so-called
$r$-modes. An interesting property of these modes that has been
pointed out by Andersson~\cite{nils98, FM98} is that these modes are
generically unstable due the Chandrasekhar-Friedman-Schutz
instability~\cite{Chandra70, FS78} and furthermore it has been
shown~\cite{AKS98, LOM98} that these modes can potentially restrict
the rotation period of  newly formed neutron stars and also that they
can radiate away detectable amounts of gravitational
radiation~\cite{Owen98}. The equations describing the perturbations of
slowly rotating relativistic stars have been derived by
Kojima~\cite{kojima92, kojima93}, and Chandrasekhar and
Ferrari~\cite{chandra91b}.


\subsection{Mode Analysis}
\label{section_4_2}

The study of stellar oscillations in a general relativistic context
already has a history of 30 years. Nevertheless, recent results have
shown remarkable features which had previously been overlooked.

Until recently most studies treated the stellar oscillations in a
nearly Newtonian manner, thus practically ignoring the dynamical
properties of the spacetime~\cite{TC67, Thorne68, PT69, Thorne69a,
  Thorne69b, lindblom83, MVS83, DL85, MVH88}. The spacetime was used
as the medium upon which the gravitational waves, produced by the
oscillating star, propagate. In this way all the families of modes
known from Newtonian theory were found for relativistic stars while in
addition the damping times due to gravitational radiation were
calculated.

Inspired by a simple but instructive model ~\cite{kokkotas86}, Kokkotas
and Schutz showed the existence of a new family of
modes: the $w$-modes~\cite{kokkotas92}. These are {\em spacetime modes} and their
properties, although different, are closer to the black hole QNMs than 
to the standard fluid stellar modes. The main characteristics of the
$w$-modes are high frequencies accompanied with very rapid
damping. Furthermore, these modes hardly excite any fluid motion. The
existence of these modes has been verified by subsequent
work~\cite{leins93, AKS95}; a part of the spectrum was found earlier
by Kojima~\cite{kojima88} and it has been shown that they exist also
for odd parity (axial) oscillations~\cite{kokkotas94}. Moreover,
sub-families of $w$-modes have been found for both the polar and axial
oscillations i.e.\ the {\em interface} modes found by Leins et
al.~\cite{leins93} (see also~\cite{AKK96}), and the {\em trapped}
modes found by Chandrasekhar and Ferrari~\cite{chandra91c} (see
also~\cite{kokkotas94, kojima95, AKK96}). Recently, it has been proven
that one can reveal all the properties of the $w$-modes even if one
``freezes'' the fluid oscillations (Inverse Cowling
Approximation)~\cite{AKS96}. In the rest of this section we shall
describe the features of both  families of oscillation modes, fluid and spacetime, for the
case $\ell=2$.


\subsubsection{Families of Fluid Modes}

For non-rotating stars the fluid modes exist {\em only for polar}
oscillations. Here we will describe the properties of the most
important modes for gravitational wave emission. These are the
fundamental, the pressure and the gravity modes; this division has
been done in a phenomenological way by Cowling~\cite{cowl41}. For an
extensive discussion of other families of fluid modes we refer the
reader to~\cite{gau95, gau96} and~\cite{MVS83, MVH88}. Tables of
frequencies and damping times of neutron star oscillations for twelve
equations of state can be found in a recent work~\cite{AK98} which
verifies and extends earlier work~\cite{lindblom83}. In Table 2 we
show characteristic frequencies and damping times of various QNM
modes for a typical neutron star. 

\begin{table}
  \begin{center}
    \begin{tabular}{|l|r|l|}
      \hline
      mode         & frequency & damping time \\
      \hline
      $f$          &  2.87 kHz & 0.11 sec \\
      $p_1$        &  6.57 kHz & 0.61 sec \\
      $g_1$        & 19.85 Hz  & years    \\
      $w_1$        & 12.84 kHz & 0.024 ms \\
      $w_{II}$     &  8.79 kHz & 0.016 ms \\
      \hline
    \end{tabular}
  \end{center}
  \caption{\it Typical values of the frequencies and the damping times
    of various families of modes for a polytropic star ($N=1$) with
    $R=8.86$\protect~km and $M=1.27 M_\odot$ are given. $p_1$ is the
    first $p$-mode, $g_1$ is the first
    $g$-mode\protect~\protect\cite{Finn86}, $w_1$ stands for the first
    {\bf curvature} mode and $w_{II}$ for the slowest damped {\bf
      interface} mode. For this stellar model there are no {\bf
      trapped} modes.}
  \label{table2}
\end{table}

\begin{itemize}
\item The $f$-mode ({\em f}undamental) is a {\em stable} mode which
  exists only for non-radial oscillations. The frequency is
  proportional to the mean density of the star and it is nearly
  independent of the details of the stellar structure. An exact
  formula for the frequency can be derived for Newtonian uniform
  density stars
  \begin{equation}
    \omega^2 = {{2\ell(\ell-1)}\over {2\ell+1}} {M\over R^3} \ .
  \end{equation}
  This relation is approximately correct also for the relativistic
  case~\cite{AKK96} (see also the discussion in
  section~\ref{section_5_4}). The $f$-mode eigenfunctions have {\em no
    nodes} inside the star, and they grow towards the surface. A
  typical neutron star has an $f$-mode with a frequency of $1.5-3$~kHz
  and the damping time of this oscillation is less than a second
  ($0.1-0.5$~sec). Detailed data for the frequencies and damping times
  (due to gravitational radiation) of the $f$-mode for various
  equations of state can be found in~\cite{lindblom83,
    AK98}. Estimates for the damping times due to viscosity can be
  found in~\cite{CL87, CLS90}.
\item The $p$-modes ({\em p}ressure or acoustic) exist for both radial
  and non-radial oscillations. There are infinitely many of them. The
  pressure is the restoring force and it experiences substantial
  fluctuations when these modes are excited. Usually, the radial
  component of the fluid displacement vector is significantly larger
  than the tangential component. The oscillations are thus nearly
  radial. The frequencies depend on the travel time of an acoustic
  wave across the star. For a neutron star the frequencies are
  typically higher than $4-7$~kHz ($p_1$-mode) and the damping times
  for the first few $p$-modes are of the order of a few seconds. Their
  frequencies and damping times  increase with the order of the
  mode. Detailed data for the frequencies and damping times (due to
  gravitational radiation) of the $p_1$-mode for various equations of
  state can be found in~\cite{AK98}.
\item The {$g$-modes} ({\em g}ravity) arise because gravity tends to
  smooth out material inhomogeneities along equipotential
  level-surfaces and buoyancy is the restoring force. The changes in
  the  pressure are very small along the star. Usually, the tangential
  components of the fluid displacement vector are dominant in the
  fluid motion. The $g$-modes require a {\em non-zero Schwarzschild
    discriminant} in order to have non-zero frequency, and if they
  exist there are infinitely many of them. If the perturbation is
  stable to convection, the $g$-modes will be stable ($\omega^2>0$);
  if unstable to convection the $g$-modes are unstable ($\omega^2<0$);
  and if marginally stable to convection, the $g$-mode frequency
  vanishes. For typical neutron stars they have frequencies smaller
  than a hundred Hz (the frequency decreases with the order of the
  mode), and they usually damp out in time much longer than a few days
  or even years. For an extensive discussion about $g$-modes in
  relativistic stars refer to~\cite{Finn86, Finn87}; and for a study
  of the instability of the $g$-modes of rotating stars to
  gravitational radiation reaction refer to~\cite{Lai98}.
\item The {$r$-modes} ({\em r}otational) in a non-rotating star are
  purely toroidal (axial) modes with vanishing frequency. In a
  rotating star, the displacement vector acquires spheroidal
  components and the frequency in the rotating frame, to first order
  in the rotational frequency $\Omega$ of the star, becomes 
  \begin{equation}
    \omega_r = {{2 m \Omega}\over {\ell (\ell+1)}}.
    \label{omega_r}
  \end{equation}
  An inertial observer measures a frequency of 
  \begin{equation}
    \omega_i = \omega_r - m\Omega.
    \label{omega_i}
  \end{equation}
  From~(\ref{omega_r}) and~(\ref{omega_i}) it can be deduced that a
  counter-rotating (with respect to the star, as defined in the
  co-rotating frame) $r$-mode appears as co-rotating with the star to
  a distant inertial observer. Thus, all $r$-modes with $\ell \geq 2$
  are generically unstable to the emission of gravitational radiation,
  due to the Chandrasekhar-Friedman-Schutz (CFS)
  mechanism~\cite{Chandra70, FS78}. The instability is active as long
  as its growth-time is shorter than the damping-time due to the
  viscosity of neutron star matter. Its effect is to slow down, within
  a year, a rapidly rotating neutron star to slow rotation rates and
  this explains why only slowly rotating pulsars are associated with
  supernova remnants~\cite{AKS98, LOM98, KS98}. This suggests that the
  $r$-mode instability might not allow millisecond pulsars to be
  formed after an accretion induced collapse of a white
  dwarf~\cite{AKS98}. It seems that millisecond pulsars can only be
  formed by the accretion induced spin-up of old, cold neutron
  stars. It is also possible that the gravitational radiation emitted
  due to this instability  by a newly formed neutron star could be
  detectable by the advanced versions of the gravitational wave
  detectors presently under construction~\cite{Owen98}. Recently,
  Andersson, Kokkotas and Stergioulas~\cite{AKSt98} have suggested
  that the $r$-instability might be responsible for stalling the
  neutron star spin-up in strongly accreting Low Mass X-ray Binaries
  (LMXBs). Additionally, they suggested that the gravitational waves
  from the neutron stars, in such LMXBs, rotating at the instability
  limit may well be detectable. This idea was also suggested  by
  Bildsten~\cite{Bild98} and studied in detail by
  Levin~\cite{Levin98}.
\end{itemize}

\begin{figure}[hptb]
  \epsfysize=9cm  
  \epsfxsize=8cm 
  \centerline{\epsffile{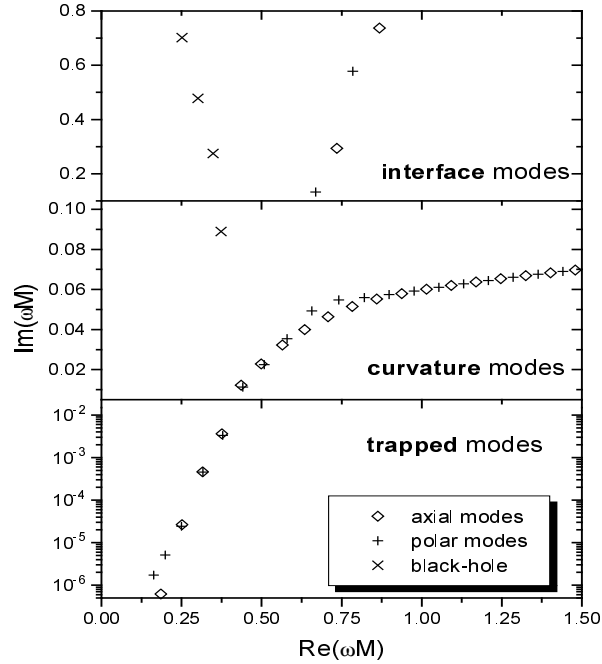}}  
  \caption{\it A graph which shows all the $w$-modes: {\bf curvature},
    {\bf trapped} and {\bf interface} both for axial and polar
    perturbations for a very compact uniform density star with
    $M/R=0.44$. The black hole spectrum is also drawn for
    comparison. As the star becomes less compact the number of {\bf
      trapped} modes decreases and for a typical neutron star
    ($M/R=0.2$) they disappear. The $Im(\omega)=1/damping$ of the {\bf
      curvature} modes increases with decreasing compactness, and for
    a typical neutron star the first curvature mode nearly coincides
    with the fundamental black hole mode. The behavior of the {\bf
      interface} modes changes slightly with the compactness. The
    similarity of the axial and polar spectra is apparent.} 
  \label{modes}   
\end{figure}   


\subsubsection{Families of Spacetime or $w$-Modes}

The spectra of the three known families of $w$-modes are different but
the spectrum of each family is similar {\em both} for polar and axial
stellar oscillations. As we have mentioned earlier they are clearly
modes of the spacetime and from numerical calculations appear to be stable.

\begin{itemize}
\item The {\bf curvature modes} are the standard
  $w$-modes~\cite{kokkotas92}. They are the most  important for
  astrophysical applications. They are clearly related to the
  spacetime curvature and exist for all relativistic stars. Their main
  characteristic is the rapid damping of the oscillations. The damping
  rate increases as the compactness of the star decreases: For nearly
  Newtonian stars (e.g.\ white dwarfs) these modes have not been
  calculated due to numerical instabilities in the various codes, but
  this case is of marginal importance due to the very fast damping
  that these modes will undergo. One of their main characteristics is
  the absence of significant fluid motion (this is a common feature
  for all families of $w$-modes). Numerical studies have indicated the
  existence of an infinite number of modes; model problems suggest
  this too~\cite{kokkotas86, BS93, amodel}. For a typical neutron star
  the frequency of the first $w$-mode is around $5-12$~kHz and
  increases with the order of the mode. Meanwhile, the typical damping
  time is of the order of a few tenths of a millisecond and decreases
  slowly with the order of the mode. 
\item The {\bf trapped modes} exist only for supercompact stars ($R
  \le 3M$) i.e.\ when the surface of the star is inside the peak of
  the gravitational field's potential barrier~\cite{chandra91c,
    kokkotas94}. Practically, the first few curvature modes become
  trapped as the star becomes more and more compact, and even the
  $f$-mode shows similar behavior~\cite{kojima95, AKK96}. The trapped
  modes, as with all the spacetime modes, do not induce any
  significant fluid motions and there are only a finite number of them
  (usually less than seven or so). The number of trapped modes
  increases as  the potential well becomes deeper, i.e.\ with
  increasing compactness of the star. Their damping is quite slow
  since the gravitational waves have to penetrate the potential
  barrier. Their frequencies can be of the order of a few hundred Hz
  to a few kHz, while their damping times can be of the order of a few
  tenths of a second. In general no realistic equations of state are
  known that would allow the formation of a sufficiently compact star
  for the trapped modes to be relevant.
\item The {\bf interface modes}~\cite{leins93} are extremely rapidly
  damped modes. It seems that there is only a finite number of such
  modes ($2-3$ modes only)~\cite{AKK96}, and they are in some ways
  similar to the modes for acoustic waves scattered off a hard
  sphere. They do not induce any significant fluid motion and their
  frequencies can be from 2 to 15 kHz for typical neutron stars while
  their damping times are of the order of less than a tenth of a
  millisecond.
\end{itemize}


\subsection{Stability}
\label{section_4_3}

The stability of radial oscillations for non-rotating stars in general
relativity is well understood. Especially, the stability of static
spherically symmetric stars can be determined by examining the
mass-radius relation for a sequence of equilibrium stellar models, see
for example Chapter 24 in~\cite{mtw}. The radial perturbations are
described by a Sturm-Liouville second order equation with the
frequency of the mode being the eigenvalue $\omega^2$, then for real
$\omega$ the modes will be stable while for imaginary $\omega$ they
will be unstable~\cite{Chandra64}, see also Chapter 17.2
in~\cite{ST83}.

The stability of the non-radially pulsating stars (Newtonian or
relativistic) is determined by the Schwarzschild discriminant 
\begin{equation}
  S(r) = {{d p}\over {d r}} - {{\Gamma_1 p}\over {\rho+p}}
  {{d \rho}\over {d r}},
\end{equation}  
where $\Gamma_1$ is the star's adiabatic index. This can be understood
if we define the local buoyancy force $f$ per unit volume acting on a
fluid element displaced a small radial distance $\delta r$ to be 
\begin{equation}
  f \sim -g(r) S(r) \delta r,
\end{equation} 
where $g$ is the local acceleration of gravity. When $S$ is negative
in some region the buoyancy force is positive and the star is unstable
against convection, while when $S$ is positive the buoyancy force is
restoring and the star is stable against convection. Another way of
discussing the stability is through the so-called Brunt-V\"ais\"al\"a
frequency $N^2=g S(r)$ which is the characteristic frequency of the
local fluid oscillations. Following earlier discussions when $N^2$ is
positive, the fluid element undergoes oscillations, while when $N^2$
is negative the fluid is locally unstable. In other words, in
Newtonian theory stability to non-radial oscillations  can be
guaranteed only if $S>0$ everywhere within the star~\cite{Cox}. In
general relativity~\cite{DI73}, this is a sufficient condition, and so
if $S>0$ the quasi-normal modes are stable. For an extensive
discussion of stellar instabilities for both non-rotating and rotating
stars  (which are actually more interesting for the gravitational wave
astronomy) refer to~\cite{BFS87, lindblom97, Nick98}.

For completeness the same applies as outlined at the end of
section~\ref{section_3_3}. A model calculation of Price and
Husain~\cite{PH92}, however indicated that the nearly Newtonian
quasi-normal modes might be a basis for the fluid
perturbations. Further mathematical investigation is needed to clarify
this issue.


\newpage

\section{Excitation and Detection of QNMs}
\label{section_5}

A critical issue related to the discussions of the previous sections
is the excitation of the QNMs. The truth is that, although the QNMs
are predicted by our perturbation equations, it is not always clear
which ones will be excited and under what initial conditions. As we
have already mentioned in the introduction there is an excellent
agreement between results obtained from perturbation theory and full
nonlinear evolutions of Einstein equations for head-on collisions of
two black holes. Still there is a degree of arbitrariness in the
definition of initial data for other types of stellar or black hole
perturbations. This due to the arbitrariness in specifying the
gravitational wave content in the initial data. 

The construction of acceptable initial data for the evolution of
perturbation equations is not a trivial task. In order to specify
astrophysically relevant initial data one should first solve the fully
nonlinear 3-dimensional initial value problem for (say) a newly formed
neutron star that settles down after core collapse or two colliding
black holes or neutron stars. Afterwards, starting from the Cauchy
data on the initial hypersurface, one can evolve forward in time with
the linear equations of perturbation theory instead of the full
nonlinear equations. Then most of the long-time evolution problems of
numerical relativity (throat stretching when black holes form,
numerical instabilities or effects due to the approximate outer
boundary conditions) are avoided. Additionally, the interpretations of
the computed fields in terms of radiation is
immediate~\cite{AP96}. This scheme has been used by  Price and
Pullin~\cite{PP94} and Abrahams and Cook~\cite{AC94} with great
success for head-on colliding black holes. The success was based on
the fact that the bulk of the radiation is generated only in the very
strong-field interactions around the time of horizon formation and the
radiation generation in the early dynamics can be practically ignored
(see also the discussion in~\cite{AST95}). The extension of this
scheme to other cases, like neutron star collisions or supernovae
collapse, is not trivial. But if one can define even numerical data on
the initial hypersurface then the perturbation method will be probably
enough or at least a very good test for the reliability of fully
numerical evolutions. For a recent attempt towards applying the above
techniques in colliding neutron stars see~\cite{ARAKLP, AKLP98}.
 
Before going into details we would like to point out an important
issue, namely the effect of the potential barrier on the QNMs of
black holes. That is, for any set of initial data that one can impose,
the QNMs will critically depend on the shape of the potential barrier,
and this is the reason that the close limit approximation of the two
black-hole collision used by Price and Pullin~\cite{PP94} was so
successful, because whatever initial data you provide inside the $r<
3M$ region (the peak of the potential barrier is around $3M$) the
barrier will ``filter'' them and an outside observer will observe only
the QNM ringing (see for example recent studies  by Allen, Camarda
and Seidel~\cite{ACS98}). This point of view is complementary to the
discussion earlier in this section, since roughly speaking even before
the creation of the final black hole the common potential barrier has
been created and anything that was to be radiated had to be
``filtered'' by this common barrier.


\subsection{Studies of Black Hole QNM Excitation} 
\label{section_5_1}

In the study of QNMs of black holes the attention, in most cases, was
focused on estimating the spectrum and its properties. But
there is limited work in the direction of understanding what details
of the perturbation determine the strength of the QNM ringing. 

In 1977 Detweiler~\cite{Det77} discussed the resonant oscillations of
a rotating black hole, and after identifying the QNMs as ``resonance
peaks'' in the emitted spectrum he showed that the modes formally
correspond to poles of a Green function to the inhomogeneous Teukolsky
equation~\cite{Teu73}. This idea has been extended in a more
mathematically rigorous way by Leaver~\cite{Leaver86}. Leaver extracts
the QNM contribution to the emitted radiation as a sum over
residues. This sum arises when the inversion contour of the Laplace
transform, which was used to separate the dependence on the spatial
variables from the time dependence, is deformed analytically in the
complex frequency plane. In this way the contribution from the QNM can
be accounted for. Sun and Price~\cite{SP88,SP90} discussed in detail
the way that QNM are excited by given Cauchy data based to some extent
on numerical results obtained by Leaver~\cite{Leaver86}. Lately,
Andersson~\cite{Nils95} used the phase-integral method to determine
some characteristics of the QNM excitation.

\begin{figure}[hptb]
  \epsfysize=7cm  
  \epsfxsize=8cm 
  \centerline{\epsffile{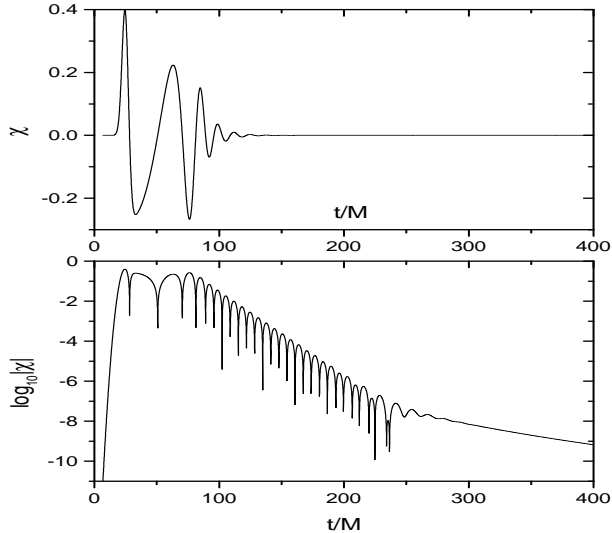}}  
  \caption{\it The response of a Schwarzschild black hole as a
    Gaussian wave packet impinges upon it. The QNM signal dominates
    the signal after $t\approx 70M$ while at later times (after
    $t\approx 300M$) the signal is dominated by a power-law fall-off
    with time.}
  \label{bhwave}   
\end{figure}


\subsection{Studies of Stellar QNM Excitation}
\label{section_5_2}

The discussion in the previous sections has shown that we have an
acceptable knowledge of stellar pulsations and that we can extract
information from the detection of such oscillations. But as was clear
from the discussion earlier in this section, the energy released as
gravitational radiation during the stellar collapse is basically
unknown. Additionally, it is not known how the energy is distributed
in the various fluid and spacetime modes. Both uncertainties depend
strongly on the details of gravitational collapse, or in general the
mechanism that excites the modes. Unless full 3D general relativistic
codes are generated for the gravitational collapse or the final stages
of binary coalescence we will never be able to give definite answers
to the above questions. In the meantime, perturbation theory is a
reliable way to get some first hints and indications. A survey of the
literature reveals several indications that the pulsation modes are
present in the gravitational wave signals from coalescing
stars. Waveforms obtained by Nakamura and Oohara~\cite{nakamura91}
show clear mode presence. Ruffert et al.~\cite{ruffert} have also
obtained gravitational wave signals from coalescing stars that show
late-time oscillations. Their waveforms and spectra show oscillations
at frequencies between 1.5 and 2 kHz.

The situation is similar for rotating core collapse. Most available
studies use Newtonian hydrodynamics and account for gravitational wave
emission through the quadrupole formula. The collapse of a
non-rotating star is expected to bounce at nuclear densities, but if
the star is rotating the collapse can also bounce at subnuclear
densities because of the centrifugal force. In each case, the emerging
gravitational waves are dominated by a burst associated with the
bounce. But the waves that follow a centrifugal bounce can also show
large amplitude oscillations that may be associated with pulsations in
the collapsed core. Such results have been obtained by M\"onchmeyer et
al.~\cite{monch91}. Some of their models show the presence of modes
with different angular dependence superimposed. Typically, these
oscillations have a period of a few ms and damp out in 20~ms. The
calculations also show that the energy in the higher multipoles is
roughly three orders of magnitude smaller than that of the
quadrupole. More recent simulations by Yamada and Sato~\cite{yamada}
and Zwerger and M\"uller~\cite{zwerger} also show post-bounce
oscillations. The cited examples are encouraging, and it seems
reasonable that besides the fluid modes the spacetime modes should
also be excited in a generic case. To show that this is the case one
must incorporate general relativity in the simulations of collapse and
coalescence. As yet there have been few attempts to do this, but an
interesting example is provided by the core collapse studies of Seidel
et al.~\cite{ed2, ed1, seidel90, seidel91}. They considered axial (odd
parity) or polar (even parity) perturbations of a time-dependent
background (that evolves according to a specified collapse
scenario). The extracted gravitational waves are dominated by a sharp
burst associated with the bounce at nuclear densities. But there are
also features that may be related to the fluid and the
$w$-modes. Especially for the axial case, since there are no axial
fluid modes for a non-rotating star, it is plausible that this mode
corresponds to one of the axial $w$-modes of the core. Furthermore,
the power spectrum for one of the simulations discussed in~\cite{ed2}
(cf.\ their Fig.\ 2) shows some enhancement around 7 kHz.

\begin{figure}[hptb]
  \epsfysize=6cm  
  \epsfxsize=7cm 
  \centerline{\epsffile{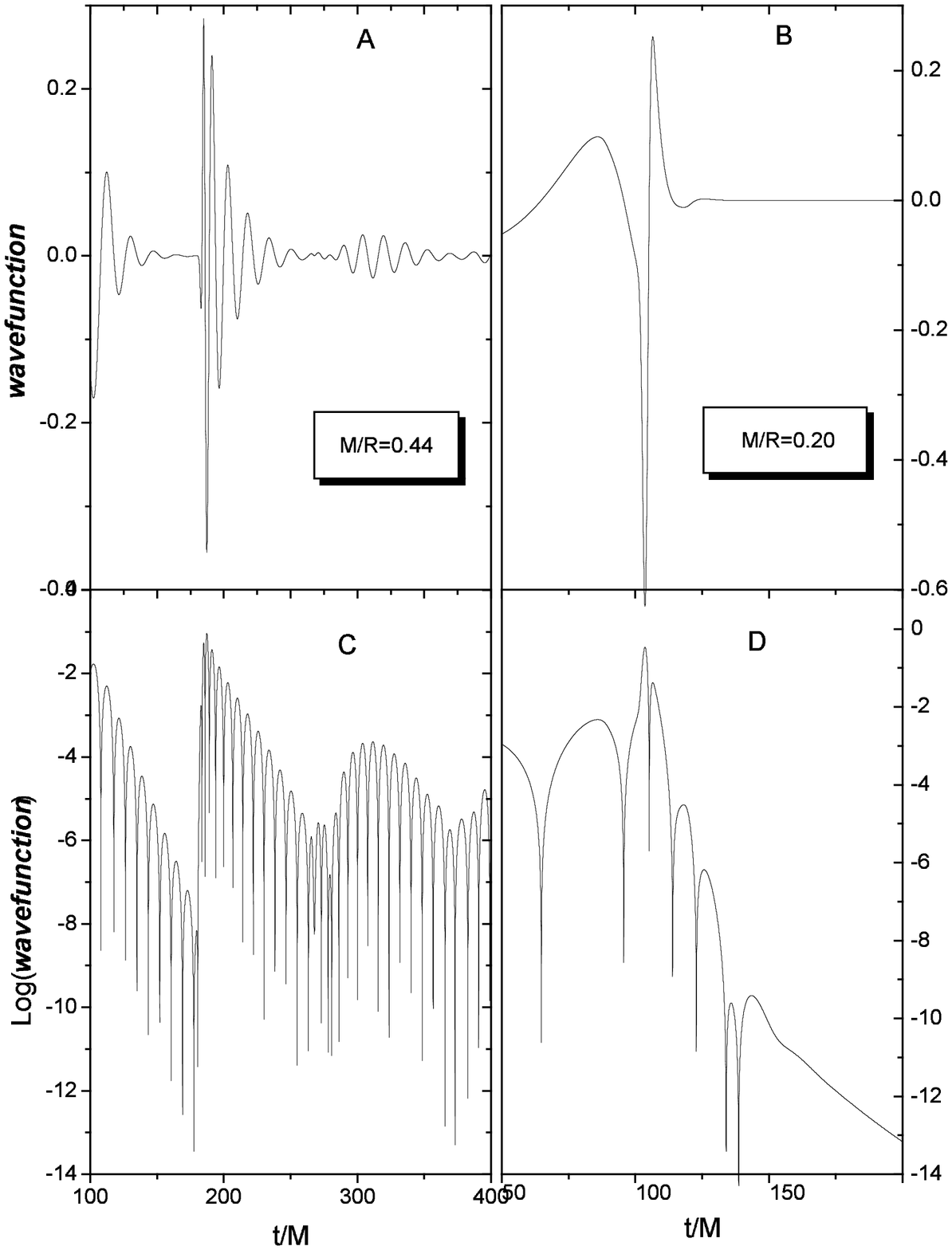}}  
  \caption{\it Time evolution of axial perturbations of a neutron star,
    here only axial $w$-modes are excited. It is apparent in panel D
    that the late time behavior is dominated by a time tail. In the
    left panels (A and C) the star is ultra compact ($M/R=0.44$) and
    one can see not only the {\bf curvature} modes but also the {\bf
      trapped} modes which damp out much slower. The stellar model for
    the panels (B and D) is a typical neutron star ($M/R=0.2$) and we
    can see only the first {\bf curvature} mode being excited.}
  \label{nswave1}
\end{figure}

Recently, Andersson and Kokkotas~\cite{ak96} have studied the
excitation of axial modes by sending gravitational wave pulses to hit
the star (see figure~\ref{nswave1}). The results of this study are
encouraging because they provide the first indication that the
$w$-modes can be excited in a dynamical scenario. Similar results have
recently been obtained by Borelli~\cite{BF96} for particles falling
onto a neutron star. The excitation of the axial QNMs  by test
particles scattered by a neutron star have been recently studied in
detail~\cite{TSM99, FGB99, AP99} and some interesting conclusions can
be drawn. For example, that the degree of excitation depends on the
details of the particle's orbit, or that in the cases that we have
strong excitation of the quadrapole oscillations there is a comparable
excitation of higher multipole modes.

These results have prompted the study of an astrophysically more
relevant problem, the excitation of even parity (or polar) stellar
oscillations. In recent work Allen et al.~\cite{aaks} have studied the
excitation of the polar modes using two sets of initial data. First,
as previously for the axial case, they excited the modes by an
incoming gravitational wave pulse. In the second case, the initial
data described an initial deformation of both the fluid and the
spacetime. In the first case the picture was similar to that of the
axial modes i.e.\ the star was excited and emitted gravitational waves
in both spacetime and the fluid modes. Nearly all of the energy was
radiated away in the $w$-modes.

This should be expected since the incoming pulse does not have enough
time to couple to the fluid (which has a much lower speed of
propagation of information). The infalling gravitational waves are
simply affected by the spacetime curvature associated with the star,
and the outgoing radiation contains an unmistakable $w$-mode
signature. In the second case one has considerable freedom in choosing
the degree of initial excitation of the fluid and the spacetime. In
the study the choice was some ``plausible'' initial data inspired by
the treatment of the problem in the Newtonian limit. Then the energy
emitted as gravitational waves was more evenly shared between the
fluid and the spacetime modes, and one could see the $f$, $p$, and $w$
modes in the signal. The characteristic signal was (as expected) a
short burst ($w$-modes) followed by a slowly damped sinusoidal wave
(fluid modes).

The previous picture emerged as well in a recent study in the close
limit of head-on collision of two neutron stars~\cite{ARAKLP}. There
the excitation of both fluid and spacetime modes is apparent but most
of the energy is radiated via the fluid modes. These new results show
the importance of general relativity for the calculation of the energy
emission in violent processes. Up to now the general way of
calculating the energy emission has been the quadrupole formula. But
this formula only accounts for the fluid deformations and motions, and
as we have seen this accounts only for a part of the total energy. 

\vskip 4cm

\begin{figure}[hptb]
  \epsfysize=6cm  
  \epsfxsize=7cm 
  \centerline{\epsffile{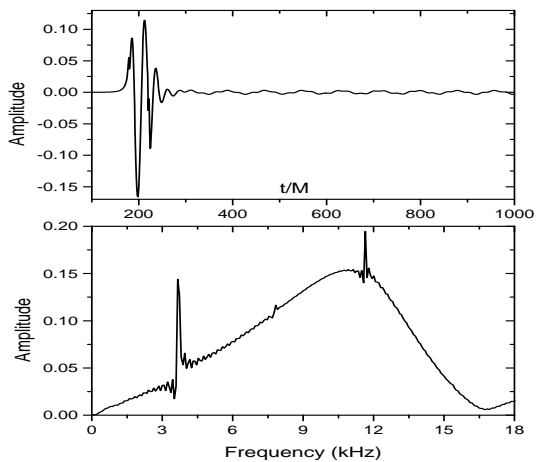}}
  \caption{\it Excitation of polar modes due to an initial deformation
    of the star. The initial burst which is dominated mainly by the
    first $w$-mode is followed by a sinusoidal waveform  dominated by
    the $f$ and the first $p$-mode. In the upper panel the actual
    waveform is shown, while in the lower panel is its power
    spectrum. The wide peak in the power spectrum corresponds to the
    $w$-mode, while the sharp peaks correspond to the various fluid
    modes.}
  \label{nswave2}   
\end{figure}


\subsection{Detection of the QNM Ringing}
\label{section_5_3}

It is well known in astrophysics that many stars will end their lives
with a violent supernova explosion. This will leave behind a compact
object which will oscillate violently in the first few seconds. Huge
amounts of gravitational radiation will be emitted and the initial
oscillations will consequently damp out. The gravitational waves will
carry away information about the compact object. If the supernova
remnant is a black hole we will observe a short monochromatic burst
lasting a few tenths of a millisecond (see figures~\ref{qnm1}
and~\ref{bhwave}). If it is a neutron star we will observe a more
complicated signal which will be the overlapping of several
frequencies (see figure~\ref{nswave2}). The stellar signal will
consist of a short burst (similar to that from a black hole) followed
by a long lasting sinusoidal wave. Supernovae are quite frequent, the
expected event rate is 5.8 $\pm$ 2.4 events per century per
galaxy~\cite{BT91}. The amount of energy emitted in such an event
depends on the details of the collapse. A spherical collapse will not
produce any gravitational waves at all, while as much as  $10^{-3}
M_\odot c^2$~\cite{bfs97} can be radiated away in a highly
non­spherical one. The collapse releases an enormous amount of energy,
at least equal to the binding energy of a neutron star, about $0.15
M_\odot c^2$. Most of this energy should be carried away by neutrinos,
and this is supported by the neutrino observations at the time of
supernova SN1987A. But even if only $1\%$ of the energy released in
neutrinos is radiated in gravitational waves then the above number
makes sense. The present numerical codes used to simulate collapse
predict that the energy emitted as gravitational waves will be of the
order of $10^{-4}-10^{-7} M_\odot c^2$~\cite{BM94}. However, most of
these codes are based on  Newtonian dynamics and the few fully general
relativistic ones are not 3-dimensional. Modern computers are still
not able to perform realistic simulations of gravitational collapse in
3D, including all the important nuclear reactions and neutrino and
photon transport. For example, most of the codes fail to explain the
high average pulsar velocity which is believed to be a result of a
boost that the neutron star gets during the collapse due to anisotropy
in the neutrino distribution~\cite{Burrows}. 

Although collapse may be the most frequent source for excitation of
black hole and stellar oscillations there are other situations in
which significant pulsations take place. For example, after the merger
of two coalescing black holes or neutron stars it is natural to expect
that the final object will oscillate. Thus the well known waveform for
inspiralling binaries~\cite{BIWW96} will be followed by a short, but
not yet properly known, period (the merger phase) and will end with
the characteristic quasi-normal signal (ringing) of the newly created
neutron star or black hole. During the inspiralling phase the stellar
oscillations can be excited by the tidal fields of the two
stars~\cite{KS95}. A detailed description of the gravitational wave
emission and detection from binary black hole coalescences can be
found in two recent articles by Flanagan and Hughes~\cite{FH98a,
  FH98b}. In the same way smaller bodies falling on a neutron star or
black hole will excite oscillations. Stellar or black hole
oscillations can also be excited by a close encounter with another compact
object~\cite{TSM99, FGB99, AP99}. 

Another potential excitation mechanism for stellar pulsation is a
starquake, e.g., associated with a pulsar glitch. The typical energy
released in this process may be of the order of $10^{-10} M_\odot
c^2$. This is an interesting possibility considering the recent
discovery of so-called magnetars: Neutron stars with extreme magnetic
fields~\cite{DT92}. These objects are sometimes seen as soft gamma-ray
repeaters, and it has been suggested that the observed gamma rays are
associated with starquakes. If this is the case, a fraction of the
total energy could be released through nonradial oscillations in the
star. As a consequence, a burst from a soft gamma-ray repeater may be
associated with a gravitational wave signal. 

Finally, a phase-transition could lead to a sudden contraction during
which a considerable part of the stars gravitational binding energy
would be released, and it seems inevitable that part of this energy
would be channeled into pulsations of the remnant. Transformation of a
neutron star into a strange star is likely to induce pulsations in a
similar fashion.

One way of calibrating the sensitivity of detectors is to calculate
the amplitude of the gravitational wave that would be produced if a
certain fraction of the released energy were converted into
gravitational waves. To obtain rough estimates for the typical
gravitational wave amplitudes from a pulsating star we use the
standard relation for the gravitational wave flux which is valid far
away from the star~\cite{bfs96}
\begin{equation}
  F = {c^3 \over 16\pi G} \vert \dot{h} \vert = 
  {1\over 4\pi r^2 } {dE \over dt},
\end{equation}
where $h$ is the gravitational wave amplitude and $r$ the distance of
the detector from the source. Combining this with i) ${dE/dt} =
E/2\tau$ where $\tau$ is the damping time of the pulsation and $E$
is the available energy, ii) the assumption that the signal is
monochromatic (with frequency $f$), and iii) the knowledge that the
effective amplitude achievable after matched filtering scales as the
square root of the number of observed cycles, $t_{\rm eff} = h\sqrt{n} 
= h\sqrt{f\tau}$, we get the estimates~\cite{ak96, AK98} 
\begin{equation}
  h_{\rm eff} \sim 2.2 \times 10^{-21} \left( {E\over 10^{-6} M_\odot c^2}
  \right)^{1/2} \left(  { 2 \mbox { kHz} \over f_{gw} } \right)^{1/2}
  \left( {50 \mbox{ kpc} \over r} \right)
  \label{sens_f}
\end{equation}
for the $f$-mode, and
\begin{equation}
  h_{\rm eff}\sim 9.7\times 10^{-22} \left( { E \over 10^{-6} M_\odot c^2}
  \right)^{1/2} \left( { 10 \mbox{ kHz} \over f_{gw} } \right)^{1/2}
  \left( { 50 \mbox{ kpc} \over r } \right)
  \label{sens_w}
\end{equation}
for the fundamental $w$-mode. Here we have used typical parameters for
the pulsation modes, and the distance scale used is that to
SN1987A. In this volume of space one would not expect to see more than
one event per ten years or so. However, the assumption that the energy
release in gravitational waves in a supernova is of the order of
$10^{-6} M_\odot c^2$ is very conservative~\cite{bfs97}.

Similar relations can be found for black holes~\cite{bfs96}:
\begin{equation}
  h_{\rm eff} \sim 5\times 10^{-22} 
  \left( {E\over 10^{-3} M_\odot c^2} \right)^{1/2} 
  \left(  { 1 \mbox{ kHz} \over f_{gw} } \right)^{1/2}
  \left( {15 \mbox{ Mpc} \over r} \right),
  \label{sens_bh1}
\end{equation}
for stellar black holes, and
\begin{equation}
  h_{\rm eff}\sim 3\times 10^{-18} 
  \left( { E \over 10^3 M_\odot c^2} \right)^{1/2} 
  \left( { 1 \mbox{ mHz} \over f_{gw} } \right)^{1/2}
  \left( { 3 \mbox{ Gpc} \over r } \right),
  \label{sens_bh2}
\end{equation}
for galactic black holes. 

An important factor for the detection of gravitational waves are the
pulsation mode frequencies. Existing resonant gravitational wave
detectors, as well as laser interferometric ones which are under
construction, are only sensitive in a certain bandwidth. The spherical
and bar detectors are typically tuned to $0.6-3$~kHz, while the
interferometers are sensitive within $10-2000$~Hz. The initial part of
the QNM waveform, which carries away whatever deformation a collapse
left in the spacetime, is expected to be for a neutron star in the
frequency range of $5-12$~kHz ($w$-mode). The subsequent part of the
waveform is constructed from combination of the $f$- and
$p$-modes. Still the present gravitational wave detectors are
sensitive only in the frequencies of the $f$-mode. For a black hole
the frequency will depend on the mass and rotation rate\footnote{For neutron stars the frequencies depend not only on the mass and rotation rate, but also on the radius and the equation of state.}, thus for a 10 solar mass black hole the frequency of
the signal will be around 1~kHz, around 100~Hz for a 100~$M_\odot$
black hole and around 1~mHz for galactic black holes.


\subsection{Parameter Estimation} 
\label{section_5_4}

For astronomy it is important not only to observe various astronomical
phenomena but also to try to mine information from these
observations. From the observations of solar and stellar oscillations
(of normal stars) astronomers have managed to get details of the
internal structure of stars. In our days the GONG program~\cite{gong}
for detailed observation of the solar seismology is well
underway. This has suggested that, in a similar way, information 
about neutron star parameters (mass, radius), and internal structure or 
the mass and the rotation rate of black holes can be found, using the theory
of QNMs. It will be instructive to briefly examine the
case of oscillating black holes since they are much ``cleaner''
objects than stars. From the normal mode analysis of black hole
oscillations we can get a spectrum which is related to the parameters
of the black hole (mass $M$ and angular momentum $a$). In particular,
for the frequency of the first quasi-normal mode (which as we have
stated previously is the most important one for the gravitational wave
detection) the following approximate relations have been
suggested~\cite{Ech89, Finn92}:
\begin{equation}
  M \omega \approx \left[ 1 - {{63} \over {100}}(1-a)^{3/10} \right]
  \approx (0.37 +0.19 a),
  \label{wkerr}
\end{equation}
\begin{equation}
  \tau \approx {{4 M}\over { (1-a)^{9/10}}} \left[ 1 - {{63} \over
  {100}}(1-a)^{3/10} \right]^{-1} \approx M(1.48+2.09 a).  
  \label{taukerr} 
\end{equation}
%
These two relations can be inverted and thus from the ``observed''
frequency and the damping time we can derive the parameters of the
oscillating black hole. In practice, the noise of the detector will
contaminate the signal but still (depending on the signal to noise
ratio) we will get a very accurate estimate of the black hole
parameters. A similar set of empirical relations cannot be derived in
the case of neutron star oscillations since the stars are not as
``clean'' as black holes, since more than one frequency
contributes. Although we expect that most of the dynamical energy
stored in the fluid oscillations will be radiated away in the
$f$-mode, some of the $p$-modes may be excited as well and a
significant amount of energy could be radiated away through these
modes~\cite{aaks}. As far as the spacetime modes are concerned we
expect that only the curvature modes (the standard $w$-modes) will be
excited, but it is possible that the radiated energy can be shared
between the first two $w$-modes~\cite{aaks}. Nevertheless, Andersson
and Kokkotas~\cite{AK98}, using the properties of the various families
of modes ($f$, $p$, and $w$), managed to create a series of empirical
relations which can provide quite accurate estimates of the mass,
radius and equation of state of the oscillating star, if the $f$ and
the first $w$-mode can be observed. In figure~\ref{ffreq} one can see
an example of the relation between the stellar parameters and the
frequencies of the $f$ and the first $w$-mode for various equations of
state and various stellar models. There it is apparent that the
relation between the $f$-mode frequencies and the mean density is
almost linear, and a linear fitting leads to the following simple
relation: 
\begin{equation}
  \omega_f (\mbox{kHz}) \approx 0.78 + 1.635 \left[ \left( {M
  \over{1.4M_\odot}} \right) \left( {{10 \mbox{ km}}\over R}
  \right)^3 \right]^{1/2}.
  \label{rfw}
\end{equation}
We can also find the following relation for the frequency of the first
$w$-mode:
\begin{equation}
  \omega_w (\mbox{kHz}) \approx \left({{10 \mbox{ km}}\over
  {R}}\right) \left[ 20.92 - 9.14 \left( {M \over {1.4
  M_\odot}}\right) \left({{10 \mbox{ km}} \over R} \right)
  \right].
  \label{rww}
\end{equation}

\begin{figure}
  \epsfxsize=6cm 
  \epsfysize=8cm 
  \centerline{\epsfbox{ffw.eps}}
  \epsfxsize=6cm
  \epsfysize=8cm
  \centerline{\epsfbox{fww.eps}}
  \caption{The upper graph shows the numerically obtained $f$-mode
    frequencies plotted as functions of the mean stellar density. In
    the second graph the functional $R \omega_w$ is plotted as a
    function of the compactness of the star ($M$ and $R$ are in km,
    $\omega_{f\mbox{\scriptsize-mode}}$ and
    $\omega_{w\mbox{\scriptsize-mode}}$ in kHz). The letters A, B, C,
    \dots\ correspond to different equations of state for which one
    can refer to\protect~\protect\cite{AK98}.}
\label{ffreq}
\end{figure}

From tests performed using polytropic stars to provide data for the
above relations it was seen that these equations  predict the masses
and the radii of the polytropes usually with an error less than
10\%. There is work underway towards extracting the parameters of the
star from a noisy signal~\cite{KAA98}.


\newpage

\section{Numerical Techniques} 
\label{section_6}

For each candidate source of gravitational waves, gravitational wave
astronomy needs answers to two questions. Firstly, how much energy
will be carried away by the emitted gravitational waves? Secondly,
which are the ``preferred'' frequencies at which a 10~$M_\odot$ black
hole or a neutron star will oscillate? The answer to the first
question is that the energy will depend on the degree of asymmetry
that the process generates, and it will depend critically on the
initial data. In the previous section we have tried to provide some
guesstimates for the energy emitted during the oscillation phase of
black holes and neutron stars. The answer to the second question is
related to the numerical solution of the perturbation equations. The
numerical schemes developed for this purpose will be described in this
section. 

Let us describe why the numerical calculation of quasi-normal mode
frequencies is delicate. Consider again the case treated in
section~\ref{section_2} of the wave equation with a potential with
compact support. We try to find a complex number $s$ with negative
real part such that the solution which is $e^{-sx}$ for large positive
$x$, is $e^{sx}$ for large negative $x$. Note that these solutions
grow exponentially with $|x|$ and therefore one has to be very careful
to make sure that there is no exponentially decaying part in the
solution. The situation becomes even more complicated if we do not
know $f^{\pm}$ explicitly because one can not characterize the correct
solution by some growth property. This is for example the case for the
Schwarzschild solution.


\subsection{Black Holes}
\label{section_6_1}

We would like to point out here that the attempts to calculate the QNM
frequencies date back to the beginning of 70s. More specifically in
their study of the black hole oscillations excited by an infalling
particle Davis et al.~\cite{DRPP71} found that the peak of the
spectrum is (for a Schwarzschild black hole) at around $M\omega =
0.32$ (geometrical units). This number is very close to the one
calculated by much more accurate methods later. In what will follow we
will describe the various methods used to calculate the QNM
frequencies.


\subsubsection{Evolving the Time Dependent Wave Equation}

This approach was actually the first one used to study the QNM
excitation by Vishveshwara~\cite{vishu} but it has only been recently
revived thanks to increased power of computers. In general, one does
not need to Fourier decompose the perturbed Einstein equations but
instead evolve them for given sets of initial data and at the end to
Fourier analyze the resulting waveform. This procedure has certain
advantages since one does not need to be so careful in considering the
appropriate boundary conditions on the horizon, at infinity, on the surface or
at the center of the star. This does not mean that these boundaries are not
important for the evolution schemes. The difference is that in the
time independent case one formulates a boundary value problem, and the
eigenvalues (quasi-normal frequencies) depend critically on the correct
conditions on the various boundaries. A major disadvantage of the
evolution schemes is that one cannot get the complete spectrum of the
QNMs neither for a star nor for a black hole. The reason is that
although any perturbation is the sum of the harmonics involved, in
practice only a few of them will be observed and in the best case one
can succeed in getting a few extra modes by ``playing'' around with
the initial data. 

To be more specific, by evolving a perturbation on a black hole
background one can get a QNM signal as the one shown in figure~\ref{bhwave}, but
in this signal the Fourier transform will show that there are present
at most two frequencies (the slowest damped ones); then by doing
``matched filtering'' of this signal we will get the right frequencies
and damping times, but then all the extra modes of the spectrum are
missing! See for example the work by Bachelot and
Motet-Bachelot~\cite{BB91, BB92}, and Krivan et al~\cite{KLP96,
  KLPA97}. This is true also for stars; in recent evolutions of axial
perturbations of stellar backgrounds Andersson and
Kokkotas~\cite{ak96} saw only a few of the $w$-modes (the ones that
damped slowest), while in similar calculations for even parity stellar
perturbations Allen et al.~\cite{aaks} have seen only the $f$-mode,
$2-3$ $p$-modes and two of the $w$-modes.

Of course with more detailed studies for various sets of initial data
one might be successful to get a few more modes, but more important
than deriving extra modes is understanding the physical situation
which generates the appropriate initial data. 

Finally, the evolutions of the time dependent perturbation equations
can be extremely useful (and probably will be the only way) for the
calculation of the QNM frequencies and waveforms for the perturbations
of the Kerr-Newman black hole and for slowly and fast rotating
relativistic stars. 


\subsubsection{Integration of the Time Independent Wave Equation}

This technique was used by Chandrasekhar and Detweiler~\cite{CD75} and
is based on the definition of QNMs given in
section~\ref{section_2}. They assumed that a QNM is a solution
corresponding to incoming waves on the horizon and outgoing at
infinity. Then by taking a series expansion of the Zerilli wave
equation~(\ref{qnmwave},~\ref{rwpot},~\ref{zerpot}) at both limits (horizon and
infinity) of the form given by~(\ref{bconditions}) they found initial
values for the numerical integration of the equation. Their
integration goes from both limits towards a common point which was set
close to the peak of the potential i.e.\ around $r = 3M$. The values
of $\omega$ for which the Wronskian of the two numerically taken
solutions vanishes are the quasi­eigensolutions of the problem. In
this way they managed to calculate the first $2-3$ QNM frequencies of
the Schwarzschild black hole for various harmonic indices. The
accuracy of the method improves for increasing $\ell$. Later,
Gunter~\cite{Gunter80} and Kokkotas with Schutz~\cite{KS88} used the
same approach to calculate the QNMs of the Reissner-Nordstr\"om black
hole. 

The approach used by Nollert and Schmidt~\cite{Nollert90, NS92} is
more elaborate and based on a better estimate of the values of the
quasi-eigenfunctions on both boundaries ($\pm \infty$); this leads to
a more accurate estimate of frequencies and one also finds frequencies
which damp extremely fast. Andersson~\cite{nils92} suggested an
alternative integration scheme. The key idea is to separate ingoing
and outgoing wave solutions by numerically calculating their analytic
continuations to a place in the complex $r$-coordinate plane where
they have comparable amplitudes. This method is extremely accurate.


\subsubsection{WKB Methods}
 
This technique, originally in the form suggested by Schutz and
Will~\cite{SW85}, based on elementary quantum mechanical arguments,
was later developed into a powerful technique with which accurate
results have been derived. The idea is that one can reduce the QNM
problem into the standard WKB treatment of scattering of waves on the
peak of the potential barrier. The simplest way to find the QNM
frequencies is to use the well known Bohr-Sommerfeld (BS) rule. Using
this rule it is possible to reproduce not only the Schutz-Will formula
(\ref{qnmsw}) but also to give a way to extend the accuracy of that
formula by taking higher order terms~\cite{IW87,GWKS90}. The classical form of the BS rule
for equations like~(\ref{qnmwave}) is
\begin{equation}
  \int_{r_A}^{r_B}\left[\omega^2 - V(r)\right]^{1/2} d r= (n+{1\over 2})\pi.
  \label{BSrule}
\end{equation}
where $r_A, r_B$ are the two roots (turning points) of $\omega^2
-V(r)=0$. A more general form can be found~\cite{Dunh32, BO78} which
is valid for complex potentials. This form can be extended to the
complex $r$ plane where the contour encircles the two turning points
which are connected by a branch cut. In this way one can calculate the
eigenfrequencies even in the case of complex potentials, as it is the
case for Kerr black holes.

This method has been used for the calculation of the eigenfrequencies
of the Schwarzschild~\cite{Iyer87}, Reissner-Nordstr\"om~\cite{KS88},
Kerr~\cite{SI90, Kokkotas91} and  Kerr-Newman~\cite{KDK93} black holes
(restricted case). In general with this approach one can calculate
quite accurately the low-lying (relatively small imaginary part) QNM
modes, but it fails to give accurate results for higher-order modes.

This WKB approach was improved considerably when the phase integral
formalism of Fr\"oman and Fr\"oman~\cite{FF65} was introduced. In a
series of papers~\cite{FFAH92, AL92, AAS93a, ANS93} the method was
developed and a great number of even extremely fast damped QNMs of the
Schwarzschild black hole have been calculated with a remarkable
accuracy. The application of the method for the calculation of QNM
frequencies of the Reissner-Nordstr\"om black hole~\cite{AAS94} has
considerably improved earlier results~\cite{KS88} for the QNMs which
damp very fast. Close to the logic of this WKB approach were the
attempts of Blome, Mashhoon and Ferrari~\cite{BM84, FM84a, FM84b} to
calculate the QNM modes using an inversion of the black hole
potential. Their method was not very accurate but stimulated future
work using semi­analytic methods for estimating QNMs.
 

\subsubsection{The Method of Continued Fractions} 

In 1985, Leaver~\cite{Leaver85} presented a very accurate method for
calculating the QNM frequencies. His  method can be applied to the
calculation of the QNMs of Schwarzschild, Kerr and, with some
modifications, Reissner-Nordstr\"om black holes~\cite{Leaver90}. This
method is very accurate also for the high-order modes.

His approach was analogous to the determination of the eigenvalues of
the H$^+$ ion developed in~\cite{BH35}. A series representation of the
solution $f^-$ is assumed to represent also $f^+$ for the value of the
quasi-normal mode frequency. For normal modes the method may work
because $f^+$ is certainly bounded at infinity. In the case of
quasi-normal modes this is not so clear because $f^+$ grows
exponentially. Nevertheless, the method works very well numerically
and was improved by Nollert~\cite{Nollert93} such that he was able to
calculate very high mode numbers (up to 100,000!). In this way he
obtained the asymptotic distributions of modes described in
(\ref{asymqnm}). An alternative way of using the recurrence relations
was suggested in~\cite{MP89}.
 
Nollert~\cite{Nollert90} explains in his PhD thesis why the method
works. As initiated by Heisenberg et al.~\cite{Heisen} he considers
potentials depending analytically on a parameter $\lambda$ such that
for $\lambda = 1$ the potential has bound states -- normal modes --
and for for $\lambda = -1$ just quasi-normal modes. This is, for
example,  the case if we multiply the Regge-Wheeler potential
(\ref{rwpot}) by $-\lambda$. Assuming that the modes depend
continuously on $\lambda$, one can try to relate normal modes to
quasi-normal modes and their methods of calculation.  

In the case of QNMs of the Kerr black hole, one has to deal in
practice with two coupled equations, one which governs the radial part
(\ref{teu1b}) and another which governs the angular dependence of the
perturbation~(\ref{teu1a}). For both of them one can construct
recurrence relations for the coefficients of the series expansion of
their solutions, and through them calculate the QNM frequencies. 

For the case of the QNMs of the Reissner-Nordstr\"om black hole, the
asymptotic form of the solutions is similar to that shown in equation
(\ref{exp_hor}) but the coefficients $a_n$ are determined via a four
term recurrence relation. This means that the nice properties of
convergence of the three term recurrence relations have been lost and
one should treat the problem with great caution. Nevertheless,
Leaver~\cite{Leaver90} has overcome this problem and showed how to
calculate the QNMs for this case. 

As a final comment on this excellent method we should point out that
it has a disadvantage compared to the WKB based methods in that it is
a purely numerical method and it cannot provide much intuition about
the properties of the QNM spectrum.


\subsection{Relativistic Stars}
\label{section_6_2}

Although the perturbation equations in the exterior of a star are
similar to those of the black-hole and the techniques described
earlier can be applied here as well, special attention must be given
to the interior of the star where the perturbation equations are more
complicated. 

For the time independent case the system of
equations~(\ref{starper1},~\ref{starper2},~\ref{starper3}) inside the star reduces to a
4th order system of ODEs~\cite{DL85}. One can then even treat it as
two coupled time independent wave equations\footnote{Chandrasekhar and
  Ferrari~\cite{chandra91a} have also reduced the time independent
  perturbation equations (using a different gauge) into a 5th order
  system which involves only the spacetime perturbations with the
  fluid perturbations being calculated via algebraic relations from
  the spacetime perturbations. It was later proven by Ipser and
  Price~\cite{IP91, PI91} that this system of ODEs can be reduced to
  the standard equations in the Regge-Wheeler gauge. That this is
  possible is apparent from equations
  (\ref{starper1},~\ref{starper2},~\ref{starper3},~\ref{starconstr}) and it is discussed
  in~\cite{aaks}}. The first equation will correspond to the fluid and
the second equation will correspond to spacetime perturbations. In
this way one can easily work in the Cowling approximation (ignore the
spacetime perturbations) if the aim is the calculation of the QNM
frequencies of the fluid modes ($f$, $p$, $g$, \dots) or the Inverse
Cowling Approximation~\cite{AKS96} (ignore the fluid perturbations) if
the interest is in $w$-modes. The integration procedure inside the
star is similar to those used for Newtonian stars and involves
numerical integration of the equations from the center towards the
surface in such a way that the perturbation functions are regular at
the center of the star and the Lagrangian variation of the pressure is
zero on the surface (for more details refer to~\cite{lindblom83,
  kokkotas92}). The integrations inside the star should provide the
values of the perturbation functions on the surface of the star where
one has to match them with the perturbations of the spacetime
described by Zerilli's equation~(\ref{qnmwave},~\ref{rwpot},~\ref{zerpot}).

In principle the integrations of the wave equation outside the star
can be treated as in the case of the black holes. Leaver's method of
continued fraction has been used in ~\cite{kojima88, leins93},
Andersson's technique of integration on the complex $r$ plane was used
in~\cite{AKS95} while a simple but effective WKB approach was used by
Kokkotas and Schutz~\cite{kokkotas92, YEF94}. 

Finally, there are a number of additional approaches used in the past
which improved our understanding of stellar oscillations in GR. In the
following paragraphs they will be discussed briefly. 
\begin{itemize}
\item {\bf Resonance Approach.} This method was developed by
  Thorne~\cite{Thorne69a}, the basic assumption being that there are
  no incoming or outgoing waves at infinity, but instead standing
  waves. Then by searching for resonances one can identify the QNM
  frequencies. The damping times can be estimated from the half-width
  of each resonance. This is a simple method and can be used for the
  calculations of the fluid QNMs. In a similar fashion Chandrasekhar
  and Ferrari~\cite{chandra91a} have calculated the QNM frequencies
  from the poles of the ratio of the amplitudes of the ingoing and
  outgoing waves. 
\item {\bf Direct Numerical Integration.} This method was used by
  Lindblom and Detweiler~\cite{lindblom83} for the calculation of the
  frequencies and damping times of the $f$-modes for various stellar
  models for thirteen different equations of state. In this case,
  after integration of the perturbation equations inside the star, one
  gets initial data for the integration of the Zerilli equation
  outside. The numerical integration is extended up to ``infinity''
  (i.e.\ at a distance where the solutions of Zerilli equations
  become approximately simple sinusoidal ingoing and outgoing waves),
  where this solution is matched with the asymptotic solutions of the
  Zerilli equations which describe ingoing and outgoing waves. The QNM
  frequencies are the ones for which the amplitude of the incoming
  waves is zero. This method is more accurate than the previous one at
  least in the calculation of the damping times as has been verified
  in~\cite{AK98}, but still is not appropriate for the calculation of
  the $w$-modes.
\item{\bf Variational Principle Approach.} Detweiler and
  Ipser~\cite{DI73} derived a variational principle for non-radial
  pulsational modes. Associated with that variational principle is a
  conservation law for the pulsational energy in the star. The time
  rate of change of that pulsational energy, as given by the
  variational principle, is equal to minus the power carried off by
  gravitational waves. This method was used widely for calculating
  the $f$~\cite{Det75a} and $g$-modes~\cite{Finn86} and in studies of
  stability~\cite{DI73, Det75b}. 
\item{\bf WKB.} This is a very simple method but quite accurate, and
  contrary to the previous three methods it can be used for the
  calculation of the QNM frequencies of the $w$-modes (this was the
  first method used for the derivation of these modes). In practice
  one substitutes the numerical solutions of the Zerilli equation with
  their approximate WKB solutions and identifies the QNM frequencies
  as the values of the frequency for which the amplitude of the
  incoming waves is zero~\cite{kokkotas92, YEF94}.
\item{\bf WKB-Numerical.} This is a combination of direct integration
  of the Zerilli equation and the WKB method. The trick is that
  instead of integrating outwards, one integrates inwards (using
  initial data at infinity for the incoming wave solution); this
  procedure is numerically more stable than the outward
  integration. On the stellar surface one needs of course not the
  solution for incoming waves but the one for outgoing
  ones. But through WKB one can derive approximately the value of the
  outgoing wave solution from the value of the incoming wave
  solution. In this way errors are introduced, nevertheless the
  results are quite accurate~\cite{kokkotas92}.
\end{itemize}


\newpage

\section{Where Are We Going?}
\label{section_7}

From the previous discussions we can draw some general conclusions and
suggestions for future work. We believe that independent of the
advancement of numerical relativity and computer power there remains
much future work in perturbation theory, in particular  parallel to
the numerical relativity efforts.


\subsection{Synergism Between Perturbation Theory and Numerical Relativity}
\label{section_7_1}

Fifteen years ago with the advancements in computer power, we could
not think that perturbation theory would continue to play such an
important role in many problems. Today it is widely accepted that
there is a need for the development of new approximation schemes to
accompany large scale simulations. Fully relativistic computer
simulations  give only numerical answers to problems; often these
answers do not provide physical understanding of which principles are
important, or even what principles govern a given process. Even more, in
some cases, simulation results can be simply incorrect or
misleading. By closely coupling various perturbation schemes it is
possible to interpret and confirm simulation results.

For example, during the late stages of black hole or stellar
coalescence or supernovae collapse, the system settles down to a
slightly perturbed black hole or neutron star. Numerical codes,
evolving the full nonlinear Einstein equations, should be able to
accurately compute the waveforms  required for gravitational wave
detection. Parallel to this, it should be possible to evolve the
perturbations of both black holes and stars (governed by their own
linear evolution equations) for the same set of initial data. This is
an important check of the fully nonlinear codes. An excellent example
is the head-on collision of two black holes. This sounds like an
impossible task for perturbation theory but it can be achieved if the
two black holes are close together and they can be considered as
having already merged into a single perturbed black hole (close
limit); see more details in a recent review by
Pullin~\cite{Pullin98}. Much work has already been performed for
head-on collisions of two non-rotating black holes but the more
realistic problem for the inspiral collision of rotating black holes
is still open.

Following in the same spirit are more recent
calculations~\cite{ARAKLP} of the close limit for two identical
neutron stars. The results show the excitation of stellar QNMs and it
remains for numerical relativity to verify the results. 

Finally, perturbation theory can be also used as a tool to construct a
gauge invariant measure of the gravitational radiation in a
numerically generated perturbed black hole spacetime~\cite{ACS98,
  Seidel98}.


\subsection{Second Order Perturbations}
\label{section_7_2}

A basic feature of linearized perturbation theory is that there is no
``built-in'' indication of how good the approximation is. But if one
has results for a physical quantity to second order in a perturbation
parameter, then the difference between the results of the first and
second order theory is a quantitative indication of the error in the
perturbation calculation. From this point of view second order
perturbation theory is a practical tool for
calculations. Nevertheless, there remain many technical problems in
this approach, for example the gauge that should be
used~\cite{GNPP98c, BMMS97} and the  amount of calculations. Thus
second order perturbation theory has turned out to be much more
difficult than linearized theory, but if one overcomes these
difficulties second order calculations will be a great deal easier
than numerical relativity. From this point of view we encourage work
in this direction. In particular, the perturbations of stellar
oscillations should be extended to second order, and the study of the
second order perturbations of black holes extended to the Kerr case.


\subsection{Mode Calculations}
\label{section_7_3}

Although the modes have been well studied there remain a few open
issues. In the black hole case, the Kerr-Newman spectrum is known only
in a restricted case~\cite{KDK93}, so the general case should be
studied, probably by evolving the perturbation equations. For
non-rotating stars, there remain certain technical questions, i.e.\ to
find if there exist a class of $w$-modes with even larger imaginary
part or the existence of extra interface modes. But for rotating
stars, even slowly rotating ones, there remains much
work~\cite{Nick98}. There should be estimates of the fluid and
spacetime modes for slowly rotating stars for various realistic
equations of state, and the results should be incorporated into the
method suggested in~\cite{AK98} for the estimates of the stellar
parameters. There should be work towards understanding the possible
interaction of $r$-modes with $g$-modes~\cite{schutz, friedman}. And
finally the time dependent perturbation equations for
rotating stars should be evolved, because in this way we expect to see most of the new features that
rotation induces in the spectra (splitting, instabilities etc).


\subsection{The Detectors}
\label{section_7_4}

As we have mentioned earlier in section~\ref{section_5}, there are
techniques available for the extraction of the QNM signal from the
noise of the detectors~\cite{Ech89, FH98a, FH98b}. But there are still
issues related to the sensitivity of the planned detectors in parts of
the spectrum. For example, the QNM frequencies of stellar black holes
($10 - 100 M_\odot$) will be of around $100-1000$~Hz, i.e.\ in the
frequencies where the laser interferometers are sensitive. The QNM
frequencies of galactic size black holes will be detectable only from
space (LISA) since their frequencies will be in the mHz regime. For
the QNM frequencies of stars, there is a lack of appropriate
detectors. Laser interferometers will be sensitive enough in the
frequency regime of the $f$-mode but it will be very hard to detect
signals in the frequencies of the $p$- and $w$-modes. Nevertheless,
from the discussion in section~\ref{section_5} it is apparent that
there is a wealth of information in the signal of oscillating neutron
stars, and in order to extract this information we need the $p$-
or/and $w$-modes. This suggests that the ideas considering detectors,
or arrays of detectors, in this high frequency regime~\cite{papa}
should be considered more seriously.


\newpage

\section{Acknowledgments}

We are grateful to G.~Allen for the suggestions and the careful
reading of the manuscript. Suggestions by N.~Andersson, H.~Beyer,
A.~Rendall and N.~Stergioulas are gratefully acknowledged.

\newpage

\section{Appendix: Schr\"odinger Equation Versus Wave Equation}
\label{appendix_1}

The purpose of this appendix is a brief comparison of properties of
the Schr\"o\-dinger equation 
\begin{equation}
  i\dot\Psi = \left(-{d^2\over {dx^2}}+ V(x)\right)\Psi.
\end{equation}
and the wave equation
\begin{equation} 
  -\ddot\Phi = \left(-{d^2\over {dx^2}}+ V(x)\right)\Phi
\end{equation}
for the same potential $V(x)$.

The ansatz of ``stationary states''
\begin{equation}
  \Psi=e^{-iEt}\psi (x)
\end{equation}
leads to the time independent Schr\"odinger equation
\begin{equation}
  -\psi''+ V\ \psi=E\psi.
  \label{a_4}
\end{equation}
The ansatz (corresponding to Laplace transformation)  
\begin{equation}
  \Phi=e^{st}\phi (x)
\end{equation}
gives for the wave equation 
\begin{equation}
  -\phi''+ V\ \phi =-s^2\phi .
  \label{a_6}
\end{equation}
The equations~(\ref{a_4}) and~(\ref{a_6}) are the same if we set $
E=-s^2$. Hence time independent scattering theory -- the theory of the
operator $-d^2 / dx^2 + V(x) $ -- does not know whether we deal with
the wave equation or the Schr\"odinger equation.  

Consider first a positive  potential of compact support. From quantum
mechanics we know that there are no bound states. This is easy to
understand: outside the support of the potential the solutions are
$\exp (\pm x \sqrt{-E})$, hence only negative values of $E$ are
possible eigenvalues. However, if we multiply~(\ref{a_4}) by $\psi$
and integrate over all $x$ we obtain a contradiction because of the
positivity of $V$. Hence the operator $-d^2 / dx^2 + V(x)$ has no
eigenfunctions and there are only scattering states, the continuous
spectrum. The operator is selfadjoint on the Hilbert space of square
integrable functions on the real line. Its resolvent, ${\bf R}_E$ is
defined for all complex $E$ outside the continuous spectrum, which
consists of the non negative real numbers. ${\bf R}_E$ is an integral
operator whose kernel is the Green function constructed in
section~\ref{section_1}. There $G(s,x,x')$ was considered as a
function of $s$. As a function of $E$, $G$ is analytic on the whole
complex plane with the exception of  real $E\leq 0$, the continuous
spectrum. In terms of $s$, the Green function -- and the resolvent --
is defined on the ``physical half space'' $Re(s)>0$.

Resonances of Schr\"odinger operators can be defined as the
poles of the analytic extension of the Green function or the
resolvent. There is a huge amount of literature on the subject, in
particular mathematical papers. A convenient starting point may be the 
proceedings of a conference on resonances in
1984~\cite{albererio}. Existence of resonances, asymptotic
distribution and also their interpretation is treated. There is in
particular the recent development of ``Geometric scattering Theory",
where the use of ``pseudo differential operators'' gives strong and
interesting results~\cite{Melrose}. We describe one such result on the
asymptotic distribution of quasi-normal mode frequencies of the
Schwarzschild spacetime in section~\ref{section_2}. It is amusing to
note that in the field of quantum mechanics the same difficulties in
defining the notion of a resonance occurred as in relativity in the
context of ``normal modes of black holes''!~\cite{Simon} is a good
reference explaining this point.  

Historically, quasi-normal modes appeared the first time in Gamow's
treatment of the $\alpha$-decay. The model he studies is a potential
with two positive square potentials. Radioactive decay is exponential
in experiments. However, even the decay in time of solutions of the
free Schr\"odinger equation is a power law decay ($1 / t$ for
1-dimensional systems), similarly for potentials with compact
support. So we face the difficulty to characterize that part of the
time evolution, in which exponential decay is a good approximation,
knowing that the final decay is polynomial! In~\cite{Skibsted} such
estimates are derived.

Let us finally consider general potentials of
compact support. If the potential well is deep enough a finite number
of negative eigenvalues $E_n$ may exist describing  bound states of
the quantum system. What are the properties of the corresponding
solutions of the wave equation? Let $\psi_n(x)$ be an eigenfunction of 
$- d^2 / dx^2 + V(x)$ with eigenvalue $E_n<0$. Then ($ E=-s^2$) 
%
\begin{equation} 
  \Phi = e^{\sqrt{-E_n}\ t}\psi_n(x)
\end{equation} 
is a solution of the wave equation. For large positive $x$ we have
\begin{equation} 
  \psi_n= e^{-\sqrt{-E_n}\ x}
\end{equation}
as the only square integrable solution of the Schr\"odinger equation
with vanishing potential. Thus the solution of the wave equation for
large $x$ is 
\begin{equation}
  \Phi = e^{\sqrt{-E_n}\ t}e^{-\sqrt{-E_n}\ x} = e^{\sqrt{-E_n}\
  (t-x)}.
\end{equation} 
This solution grows exponentially in time and falls off exponentially
in space for $x \to\infty$. This apparently strange behaviour is
possible because the energy density of the conserved energy of the
wave equation 
\begin{equation}
  \epsilon={1\over 2}\left((\dot\Phi )^2+(\Phi')^2 +V\Phi^2\right) 
\end{equation}
is not positive definite, if the potential is somewhere negative. In
this situation the solution can grow in time and nevertheless have
conserved finite energy.

In the complex $s$ plane the eigenvalues appear as poles of the Green
function with values $s_n=\sqrt{-E_n}$. To the right of the largest
eigenvalue we have analyticity of the Green function.

Let us close this section by remarking that functional analysis
techniques can also be used to develop existence theory: In the case
discussed above, $- d^2 / dx^2 + V(x) $ is selfadjoint on the space of
square-integrable function. ``Functional calculus" can be used to show
the existence and uniqueness of solutions of the time dependent
Schr\"odinger and wave equation, given appropriate initial data. From
this point of view the time independent Green function is the primary
object, which is unique because there is a unique selfadjoint operator
$- d^2 / dx^2 + V(x) $ on the real line.


\newpage

\bibliography{article}


\end{document}